# CD44 alternative splicing is a sensor of intragenic DNA methylation in tumors

Eric Batsché[1,2,*], Oriane Mauger[1,2,3], Etienne Kornobis[1,2], Benjamin Hopkins[1, 2, 4], Charlotte Hanmer-Lloyd[1, 2, 4], Christian Muchardt[1,2]

[1] Unité de Régulation Epigénétique, Institut Pasteur, Paris, France.
[2] UMR3738, CNRS, Paris, France
[3] Ecole Doctorale Complexite du Vivant (ED515), Sorbonne Universite, Paris, France
[4] Keele University, Keele, Staffordshire, ST5 5BG United Kingdom
* corresponding author. Tel: +33 144389480; E-mail: eric.batsche@pasteur.fr

## *ABSTRACT*

DNA methylation (meDNA) is a suspected modulator of alternative splicing, while splicing in turn is involved in tumour formations nearly as frequently as DNA mutations. Yet, the impact of meDNA on tumorigenesis via its effect on splicing has not been thoroughly explored. Here, we find that HCT116 colon carcinoma cells inactivated for the DNA methylases DNMT1 and DNMT3b undergo a partial epithelial to mesenchymal transition (EMT) associated with alternative splicing of the CD44 transmembrane receptor. The skipping of CD44 variant exons is in part explained by altered expression or splicing of splicing and chromatin factors. A direct effect of meDNA on alternative splicing was sustained by transient depletion of DNMT1 and the methyl-binding genes MBD1, MBD2, and MBD3. Yet, local changes in intragenic meDNA also altered recruitment of MBD1 protein and of the chromatin factor HP1γ known to alter transcriptional pausing and alternative splicing decisions. We further tested if meDNA level has sufficiently strong direct impact on the outcome of alternative splicing to have a predictive value in the MCF10A model for breast cancer progression and in patients with acute lymphoblastic leukemia (B ALL). We found that a small number of differentially spliced genes mostly involved in splicing and signal transduction is systematically correlated with local meDNA. Altogether, our observations suggest that, although DNA methylation has multiple avenues to alternative splicing, its indirect effect may be also mediated through alternative splicing isoforms of these sensors of meDNA.

## *INTRODUCTION*

DNA methylation involves the addition of a methyl group at position 5 of cytosines (5mC) by a small family of DNA cytosine-5 methyltransferase enzymes (DNMT), which transfer methyl groups from the co-factor S-adenosyl-L-methionine to DNA (Robertson et al., 1999). This heritable epigenetic modification, crucial for mammalian development and cell differentiation, occurs predominantly at CpG dinucleotide sequences in mammals (Reik, 2007; Lister et al., 2009; Cedar and Bergman, 2012).
DNA methylation (meDNA) patterns are established during development by DNMT3a/b and maintained in differentiated cells mainly by DNMT1, which ensures CpG-specific propagation in the newly synthesized strand by recognizing hemimethylated DNA (Gowher and Jeltsch, 2018; Walton et al., 2011). The 5mC levels at individual CpG methylation results of an equilibrium between methylation and demethylation that can occur either passively or actively through hydroxymethylases enzymes (TET1/2/3). DNA methylation has mostly been described for its role in inhibition of transcriptional initiation, particularly in the context of imprinting, X-inactivation, retrotransposon silencing and at promoters containing CpG-rich regions. The majority of CpG in the human genome is methylated with the exception of those located in active promoters, enhancers and insulators. Within the body of genes, meDNA-enrichment prevents spurious RNA-polymerase II (RNAPII) entry, and cryptic transcriptional initiation (Hahn et al., 2011; Xie et al., 2011; Neri et al., 2017), while also regulating the activity of intragenic enhancers (Blattler et al., 2014; Lay et al., 2015).

In eukaryotes, pre-RNAs undergo splicing, a process of maturation by which large intervening sequences (introns) are removed, leaving a mature transcript composed only of exons spliced together. This process serves as a crucial regulatory step of gene expression and transcriptome diversification, as almost all genes produce alternative transcripts with differing exon compositions (Pan et al., 2008; Wang et al., 2008). The exon-exon junction assembly results from coordinated actions of the spliceosomal complexes and highly regulated steps of hundred of splicing factor recruitment. Splicing occurs while the genes are transcribed by the RNA polymerase II (RNAPII). Co-transcriptional splicing favors the coupling between the RNA maturation and the chromatin template that can impose variations in the RNAPII elongation kinetic and differential splicing factor recruitment through interaction with chromatin readers. These mechanisms can both influence the recognition of splice sites and modulate alternative splicing decisions (Braunschweig et al., 2013; Dujardin et al., 2014; Kornblihtt, 2015).

Several reports have demonstrated that specific modifications of histone tails plays a role in alternative splicing (Naftelberg et al., 2015). In parallel, gene body enriched meDNA is positively associated with active transcriptional elongation (Lister et al., 2009) and there is a positive correlation between the level of 5mC and exon inclusion in mRNA. Inversely, introns, pseudoexons and intronless genes exhibit weaker levels of 5mC (Lyko et al., 2010; Choi, 2010; Gelfman et al., 2013; Maunakea et al., 2013). These findings have suggested that DNA methylation plays a role in exon recognition by the spliceosome.

Furthermore, there is a complex crosstalk between DNA methylation and the histone code. For instance tri-methylation of histone H3 on lysine 36 (H3K36me3) brought by the elongative RNAPII is enriched at exons along with DNA methylation (Kolasinska-Zwierz et al., 2009; Gelfman et al., 2013). This was shown to be bound by DNMT3b in gene bodies (Baubec et al., 2015; Neri et al., 2017). Currently, the impact of H3K36me3 on alternative splicing is relatively well documented with evidences for an influence on recruitment of splicing factors (Luco et al., 2010; Pradeepa et al., 2012) and a dependency to splicing activity (De Almeida et al., 2011; Kim et al., 2011a).

DNA methylation is also linked to H3K9me3 through a complex combination of interactions between the lysine-methyl-transferases (KMT) EHMT2 and SUV39H1, and the HP1 proteins (Fuks et al., 2003; Lehnertz et al., 2003; Smallwood et al., 2007) that are involved in the modulation of the RNAPII elongation speed and alternative splicing (Saint-André et al., 2011; Ameyar-Zazoua et al., 2012; Yearim et al., 2015). However, DNA methylation and H3K9me3/HP1 are not always correlated, particularly in gene bodies with low CpG density (Hahn et al., 2011; Hon et al., 2012) . Thus, it is still unclear how HP1 contributes to the influence of DNA methylation on the alternative splicing regulation.

The direct influence of DNA methylation on alternative splicing decisions has mainly been investigated either by studying *in vitro* methylated minigene reporters integrated into the chromatin (Yearim et al., 2015), or by targeting TET DNA demethylases to highly methylated CpG-rich exons (Shayevitch et al., 2018). These approaches are designed to follow the output of splicing without modifying the cellular context, and strongly suggest that DNA methylation affects splicing decisions. Other observations have suggested a reciprocal effect of splicing on meDNA by recruitment of hydroxymethylases via splicing factors (Zheng et al., 2017). Such a mechanism will however require further investigation as a study on integrated reporter genes concludes that DNA methylation remains unmodified when splicing changes (Nanan et al., 2017). The mechanisms behind the impact of meDNA on splicing largely rely on methyl binding proteins including CTCF, MeCP2 and CTCFL(/BORIS) (Maunakea et al., 2013; Shukla et al., 2011; Singh et al., 2017; Wong et al., 2017; Zheng et al., 2017) that may assist the recruitment of splicing factors to pre-mRNA while it is transcribed. Methyl-binding-domain (MBD1 to 4) family members, frequently mutated in cancers, have, to our knowledge, never been associated with alternative splicing (Baubec et al., 2013; Du et al., 2015).

Human malignant tumors are characterized by pervasive changes in DNA methylation patterns (Bergman and Cedar, 2013). These changes include global hypomethylation in tumor cell genomes and focal hypermethylation of numerous CpG islands (Hon et al., 2012; Wild and Flanagan, 2010). Differential CpG methylation also occurs within the body of genes, although the impact of these methylation changes has not yet been clearly characterized. The link between DNA methylation and splicing raises the interesting possibility that modified DNA methylation may affect cancer progression not only by interfering with the activity of promoters, but also by generating a bias in the outcome of alternative splicing.
Aberrant splicing is frequently observed in human tumors and is usually explained by modified expression of splicing factors (Urbanski et al., 2018; Wong et al., 2018). For example, PRPF6, a component of the tri-snRNP complex, is overexpressed in a subset of primary and metastatic colon cancers, and its depletion by RNAi in cell lines reduces cell growth and decreases the production of the oncogenic ZAK kinase splice variant (Adler et al., 2014). Other examples include the role of SRSF6 and SRSF10 in colon cancers, and that of SRSF1 in breast cancer (Cohen-Eliav et al., 2013; Zhou et al., 2014; Anczuków et al., 2015). Changes in alternative splicing have been particularly well studied during epithelial to mesenchymal transition (EMT) (Pradella et al., 2017). EMT is a developmental program underlying the acquisition of mesenchymal properties by epithelial cells. This process, also linked to DNA methylation variations (Lee et al., 2016a; Pistore et al., 2017) is fundamental during embryogenesis, when regulated migration of restricted population of cells is required for organogenesis. However, it is also reactivated by cancer cells to invade adjacent tissues and to disseminate toward distant organs, representing essential steps during the progression of epithelial cancers to more aggressive stages. Genes differentially spliced during EMT programs are associated with migration and invasion (FGFR2, RON and CD44), polarity and cytoskeleton organization (NUMB, RAC and p120) and transcription regulation (TCF4/TCF7L2) (Pradella et al., 2017). In the case of CD44, normal EMT is associated with a switch from a long epithelial isoforms (CD44v) to a shorter CD44s is considered as having a causative impact on EMT. This switch results from the skipping of a series of alternative exons encoding regulatory regions involved in interactions between this cell surface glycoprotein and components of the extra-cellular matrix. This is frequently correlated with decreased expression of the splicing factor ESRP1 associated with EMT (Warzecha et al., 2009), but several other splicing factors are known to have an impact on CD44 exon inclusion, such as SAM68 responding to external cues (Matter et al., 2002; Valacca et al., 2010).

Here, we have explored the possibility of an impact of DNA methylation on alternative splicing favoring tumor progression. We show that inactivation of DNMT1 and DNMT3b in HCT116 activates multiple markers of EMT, including production of CD44s isoform skipping the variant exons. The EMT was associated with modified expression and splicing of several regulators of CD44 splicing, providing a possible explanation for the modified splicing of CD44. Yet, we also noted changes in the chromatin structure within the body of the CD44 gene, including a decreased accumulation of MBD1 protein accompanied by a large decrease in the recruitment of HP1γ , suggesting a direct link between DNA methylation and CD44 splicing. This was confirmed in HeLa cells where depletion of DNMT1 was sufficient to cause reduced levels of DNA methylation inside the body of CD44 and reduced usage of CD44 variant exons, without affecting splicing factors regulating CD44. Likewise, the CD44 variant exons were also skipped upon depletion of MBD1, MBD2, and MBD3.

Examination of a model for breast cancer tumor progression and of a cohort of patients with acute lymphoblastic leukemia (B ALL), further substantiated the correlation between the level of intragenic CD44 methylation and that of CD44 variant intron inclusion. Several genes encoding RNA binding proteins, extra-cellular matrix (ECM) binding proteins, or proteins involved in signaling, were also found differentially spliced according to local variations in DNA methylation as evidenced by different cellular models.

Altogether, our observations suggest that DNA methylation data may prove informative when evaluating the likelihood of producing splice variants of few genes involved in tumor progression which can play a role as direct sensor of the DNA methylation.

## *RESULTS*

### Inactivation of DNMT1 and DNMT3b in HCT116 affects epithelial differentiation

As a first approach to DNA methylation-guided alternative splicing and its possible impact on cancer, we examined HCT116 human colorectal cancer cells having undergone genetic disruption of the DNMT1 and DNMT3b DNA methylases. These extensively studied cells, known as DKO cells, are essentially depleted of genomic DNA methylation (Sup. Fig. S1A and S1B, and (Rhee et al., 2002)).

Observation of these cells in phase-contrast microscopy showed that they had largely lost the cobblestone shape of the original HCT116, while having acquired an elongated spindle shape (Fig. 1A, phase contrast). This fibroblastoid appearance was highly evocative of the HCT116 having initiated EMT upon inactivation of the two DNA methylase activities.

Consistent with this, Immunofluorescent staining of the cells reveals a loss of the E-cadherin (CDH1) epithelial marker, as well as a dramatic increase of the mesenchymal N-cadherin (CDH2) in the DKO (Fig. 1A). Western blot analysis confirmed the quantitative decrease of E-cadherin and also revealed increased accumulation of Vimentin (VIM) in the DKO cells (Fig. 1A).

To ensure that the mesenchymal drift of the DKO cells was not a recent consequence of their culture condition, we reanalyzed five existing sets of publicly available RNAseq data of DKO HCT116 cells (Sup. Fig. S1C, S1D, and S1E). This analysis confirmed increased expression of several mesenchymal markers in DKO cells compared to the parental HCT116, including N-cadherin (CDH2) and also TWIST1/2, ZEB2, LEF1, and Vimentin. In contrast, expression of the epithelial markers TCF4 (TCF7L3), AP2α, ESRP1 (RBM35a), RBFOX2, EPCAM, DSP (Desmoplakin) was reduced (Fig. 1B, Sup. Fig. S1E) (Batsché et al., 1998; Goossens et al., 2017; Pradella et al., 2017). Additional regulators or markers of EMT were up-regulated such as TGFβ2, WNT5B and WNT11, while WNT16 and FGFR2 were repressed (Table S1).
EMT is associated with multiple alternative splicing events, including a transition from a CD44v including alternative exons to a shorter CD44s lacking domains encoded by these exons (Fig. 1C). This decreased inclusion of variant exons in the mature CD44 mRNA was readily observed by RT-qPCR comparing DKO to the parental HCT116 cells, while expression of constant exons (C2 to C5) was comparable (Fig. 1D and Sup. Fig. S1E). Monitoring this decrease over five days of culture further demonstrated that is was independent of cell proliferation and confluency (Fig. 1E). A reduction in the complexity of the pool of CD44 isoform was also visualized by Western blots using two different antibodies, either detecting all forms of CD44 (Hermes3) or specifically recognizing an epitope on the v5 exon (Fig. 1F).
We also tested the effect of PMA, a phorbol ester activating the PKC pathway and causing increased inclusion of CD44 v5 variant exons (Matter et al., 2002), and of other exons by promoting a slowing-down of the elongating RNAPII (Batsché et al., 2006). In the DKO cells as in WT cells, the PMA treatment resulted in increased inclusion of variant exons, indicating that loss of DNA methylation does not *stricto sensu* interfere with regulation of alternative splicing by external cues (Fig. 1D). Yet, in the DKO cells the global level of variant exon inclusion remained below that observed in WT cells and this was confirmed by the detection of CD44v isoforms by Western blots in DKO cells (Fig. 1F). Decreased levels of the longer isoforms of CD44 (containing the v7v8 exons) was also observed upon Immunofluorescent staining of the DKO cells with the anti-CD44v antibody (Fig. 1A).
Finally, modified levels in CD44 global expression and exon skipping between DKO and WT cells was confirmed when examining RNA-seq from earlier studies (Fig. 1G and Sup. Fig. S1E)
The meta-analysis of these RNA-seq data further revealed that several genes important for EMT other than CD44 were differentially spliced in DKO cells, including CTNNB1 (b-catenin) and ECT2 (Table S1 and S2A).
In conclusion, we demonstrate that DNMT1/3b-depleted HCT116 cells, widely used to investigate the consequences of DNA methylation-loss on gene expression, undergo a clear loss of epithelial differentiation. This process is correlated with the skipping of CD44 variant exons while the responsiveness to the PMA is conserved. These profound changes in differentiation status may blur or hide potential direct effects of DNA methylation on alternative splicing decisions.

## Inactivation of DNMT1 and DNMT3b in HCT116 results in modified expression and splicing of a wide range of regulators of transcription and splicing.

To gain insights in the mechanisms linking loss of DNA methylation to modified alternative splicing of CD44, we first examined the impact of DNMT1 and DNMT3b inactivation on expression of known regulators of CD44 splicing. For that, we analyzed publically available

RNA-seq data (Simmer et al., 2012; Schrijver et al., 2013; Maunakea et al., 2013; Blattler et al., 2014). Charts for genes of interest are shown Sup. Fig. S2B, while RT-qPCR validations experiments conducted over a period of five days to avoid influence of cell proliferation and confluency, are shown Figure 2B.

Inactivation of the two DNMTs caused extensive reprogramming of the transcriptome and resulted mostly in gene activation as expected from the reduced promoter methylation (Sup. Fig. S1C, S1D and Fig. 2A). Transcriptome variations also originated from differentially spliced genes since 653 genes produced variant transcripts (Fig. 2A and Table S2). Note that, a very small proportion of these genes was also differentially expressed (56 were increased and 52 decreased), indicating that variant mRNAs induced by the loss of DNA methylation were rarely a consequence of modified promoter activity and/or were not targeted by the non-sense mediated decay (NMD) pathway.

When examining a series of 300 genes encoding RNA binding proteins with described activity in splicing regulation, we observed that 48 were modified either in their expression or in their splicing (Fig. 2A right panel and Tables S2B). This indicated that, in the DKO cells, the splicing machinery is subject to numerous changes. When specifically examining splicing factors known to affect exon composition of the CD44 message, we particularly noted down-regulation of ESRP1 (Fig. 1B, 2B, 2C), a splicing factor previously reported as repressed during EMT (Warzecha et al., 2009; Yae et al., 2012). Likewise, RBFOX2 and NBML1, known to affect CD44 splicing were also down-regulated in both DKO cells and during EMT (Mallinjoud et al., 2013; Venables et al., 2013; Braeutigam et al., 2014) . In contrast, many other splicing factors affecting CD44 splicing and associated with-EMT (reviewed in (Pradella et al., 2017)) were not significantly affected in their expression in the mutant. These included SAM68 (Matter et al., 2002; Batsché et al., 2006), SR proteins (SRSF1, SRSF3, SRSF8, SRSF10) (Galiana-Arnoux et al., 2003; Watermann et al., 2006; Kim et al., 2011b), and hnRNPs (hnRNPA1, hnRNPM) (Xu et al., 2014; Loh et al., 2015a, 2015b) (Fig. 2B, 2C; Sup. Fig. S2A, Tables S2A). Finally, TRA2B (Takeo et al., 2009) was not affected in its expression, but was found differentially spliced (Tables S2A, S2B; and Fig. 5E)

In addition to splicing factors, alternative splicing was shown to be affect locally – at the gene body locus - through chromatin marks and their associated readers. We therefore investigated whether some of the epigenetic modifications were altered in DKO cells.

We first examined the expression of chromatin modifiers in the DKO cells. Among these, we noted a clear decrease in both SUV39H1 expression and protein accumulation (Fig. 2B, 2C and Sup. Fig. S2A). In parallel, there was a noted increased expression of AGO1 transcripts (Fig. 2B and Sup. Fig. S2A), a component of the RNAi machinery involved in recruiting H3K9 methylase to the coding region of CD44 (Ameyar-Zazoua et al., 2012). Other factors of the PIWI family, including PIWIL3 and PIWIL4 (Table S1) were also increased in DKO cells. In addition, we also identified chromatin-modifiers affected in their splicing but not in their overall expression (Table S2A). For instance, cassette exon 10 of *EHMT2*, which does not modify the catalytic activity of the enzyme (Mauger et al., 2015), but affects its nuclear localization (Fiszbein et al., 2016) was less included in DKO cells, as visualized both at the RNA level by RT-PCR and at the protein level by western blot (Fig. 2D). This splicing change may participate also to the epithelial differentiation since EHMT2 interacting with SNAI1 favors the E-cadherin transcription (Dong et al., 2012)

Examination of ChIP-Seq data from DKO cells and parental HCT116 cells (Lay et al., 2015; Maurano et al., 2015) indicates that in the DKO cells, inside the CD44 coding region, decreased DNA methylation resulted in changes in a series of histone marks in the vicinity of exons, and that no CTCF binding site was unveiled (Sup. Fig. S2B). Consistent with changes of chromatin regulators, ChIP assays revealed reduced levels of H3K9m3 on CD44 variant exons v8, v9, v10 and less importantly on v3, v6 and v7 (Fig. 2E), as well as reduced recruitment of HP1γ on essentially the entire length of the CD44 gene body (Fig. 2E). This is locus-specific change since the overall levels H3K9me3 and HP1 proteins were unaffected in the DKO cells (Fig. 2C). We noted that the H3K9me3 (and HP1γ) distribution in the CD44 gene body in HCT116 cells is really equivalent to what it has been described previously in HeLa cells (Saint-André et al., 2011; Ameyar-Zazoua et al., 2012).

Finally, we examined the fate of proteins containing methyl-CpG-binding domains (MBD), primary candidates for the readout of DNA methylation events and a likely recruiter of chromatin remodelers or modifiers. ChIP assays using a validated antibody documented the expected reduction in the recruitment of MBD1 as a logical consequence of the reduced levels of DNA methylation (Fig. 2E), while the expression was not impaired (Fig. 2B). We did however note that MBD1 is subject to alternative splicing of one of its carboxyl-terminal exons (Table S2A).

Altogether, these data highlight an extensive reprogramming of the splicing machinery in the DKO. This supports that changes in CD44 alternative splicing are caused by a complex combination of modified expression and alternative splicing of splicing factors. Furthermore, our results suggest a direct effect of DNA methylation via the MBD1 protein and the HP1γ/H3K9me3 axis.

## Evidences for a direct effects of DNA methylation on alternative splicing upon depletion of DNMT1

To further explore the potential direct effect of DNA methylation on alternative splicing in a simplified model, we turned to short-term depletion of individual DNA, thereby overcoming transcriptomic changes linked to adaptation to a permanent loss of the DNA methylases. For that, we depleted DNMT1 in HeLa cells using two different siRNAs independently. As expected, both DNMT1 mRNAs and proteins were dramatically reduced and expression of the imprinted gene H19 was increased (Fig. 3A and Sup. Fig. S3B). Interestingly, depletion of DNMT1 resulted in significant skipping of CD44 variant exons, alike what we observed in DKO cells (Fig. 3B). This decrease in CD44 variant exon inclusion was also observed when DNMT1 was depleted using a commercially available pool of three siRNA. Consistent with our observation in DKO cells (Fig. 1D), stimulation of the cells with PMA did not compensate for the decreased inclusion of the variant exons (right panel of the Fig. 3B). Note that the decreased inclusion was not correlated with modified expression of the Sam68 splicing factor (Fig. 3A).

To correlate the decrease in variant exon inclusion with CD44 gene body DNA methylation, we carefully monitored 5mC levels by MeDIP (methylated DNA immunoprecipitation). As expected, the DNMT1 depletion promoted a significant decrease in DNA methylation inside the body of the CD44 gene, especially in the regions with highest levels of 5mC (note changes at C2, C5, and v4 (Fig. 3C). In addition, we noticed that the absolute levels of 5mC were not correlated with local CpG contents of the DNA, as for instance, on v4 and C3 which exhibit 5mC differences, while their CpG content was similar (Fig. 3C and Sup. Fig. S3C, arrows). This suggest some specificity in the selection of loci where DNMT1 maintains DNA methylation inside the CD44 gene, a notion compatible with DNA methylation affecting co-transcriptional events such as splicing.

We next used the MeDIP approach to explore a possible feedback from splicing to the DNA methylation. Indeed, this phenomenon was documented for H3K36me3 (De Almeida et al., 2011; Kim et al., 2011a), a histone mark involved in recruiting the de novo DNA methylase DNMT3b (Baubec et al., 2015; Neri et al., 2017), suggesting a potential effect of alternative splicing on the 5mC level of variant exons. Here, we took advantage of the increased inclusion of CD44 alternative exons upon stimulation of the cells with PMA. In these experiments, PMA has been added for short-time (2h) and a longer time (24h) periods to evaluate a potential delay in the establishment of DNA methylation change (Fig. 3D). We did not observe any change in intragenic CD44 DNA methylation upon PMA treatment, indicating that the equilibrium between *de novo* DNA methylation and demethylation is not modified by the regulation of alternative splicing. This in accordance with the lack of inhibitory effect on the variant exon level of the depletion of the *de novo* DNA methylases DNMT3a or DNMT3b (Sup. Fig. S3D and S3E). In contrary DNMT3a caused increased inclusion of several CD44 exons, but the reason of this effect will require further investigation.

A direct effect of DNA methylation on alternative splicing would predict DNA methylases and of methyl-binding proteins to have similar effects on exon inclusion. To test this hypothesis, we depleted HeLa cells from MeCP2, MBD1, MBD2, MBD3, and MBD4 using pools of siRNAs. These methyl-binding proteins are all expressed in HeLa cells (Sup. Fig. S3F) and, when antibodies were available, we detected a partial localization to actively transcribed chromatin (Sup. Fig. S3G). This is consistent with a role of MBD proteins beside transcription inhibition. Alike DNMT1, depletion of MBD1, MBD2, or MBD3 had a very clear inhibitory effect on inclusion of CD44 alternative exons (Fig. 3E and Sup. Fig. 3H and 3I). These observations indicate that the outcome of CD44 pre-mRNA splicing is affected by the availability of methyl-binding proteins. This supports a direct effect of DNA methylation maintenance by DNMT1 within CD44 gene body, possibly establishing a chromatin template compatible with an optimal action of splicing factors and perpetuating the production of specific CD44 splice isoforms.

To further support a direct role of DNA methylation on CD44 splicing regulation, we explored whether a short-term depletion of DNMT1 would affect expression of splicing regulators. To address this question, we analyzed effect of DNMT1 depletion using independently the 2 siRNAs previously used on gene expression (Sup. Fig. S3A) with a transcriptome-wide approach (exon-array). This approach confirmed the increased expression of H19 in the absence of DNMT1. However, the overall impact of the DNMT1-depletion on the transcriptome was very limited, with only 117 genes affected by both siRNAs at the level of their transcription (Sup. Fig. S3J and Table S3A). Remarkably, among the 300 annotated splicing factors, only Peptidylprolyl Isomerase H (PPIH), that mediate interactions between spliceosomal proteins, was downregulated 1.5-fold. The arrays confirmed that ESRP1 was not expressed in HeLa cells. None of the known regulators of CD44 alternative splicing were deregulated, neither at the expression nor at the splicing level.
Thus, DNMT1-depletion appeared as an opportunity to explore direct effects of DNA methylation on alternative splicing in the absence of any major deregulation of splicing-factor expression. Examination of the exon array at the level of splicing revealed that 12 genes were affected by both siRNAs (Sup. Fig. S3 K, L). Several of these genes are related to cancer, including CD44, as well as DST, GLS, GNAS, KIF1B and AHRR.
Altogether, this argued in favor of a direct effect of DNA methylation at a small number of genes, and suggested that maintenance of DNA methylation by DNMT1 is necessary to stabilize the production of specific splice isoforms.

## Correlation between DNA methylation and alternative splicing at a subset of splicing regulators in patients with acute lymphocytic leukemia.

The observations described above suggested that splicing of a subset of genes is regulated by DNA methylation. With the objective of validating this observation in the context of cancer, we examined a cohort of pediatric patients with acute lymphocytic leukemia (B ALL) for which both methylome and transcriptome data were available with their normal counterparts as controls *i.e*. healthy precursor B-cells isolated from umbilical cord blood (HBC) (Almamun et al., 2015).
The methylome of these cells was mapped using methylated-CpG island recovery assay (MIRA) which uses the methyl binding activity of MBD2B and MBD3L1 to enrich in methylated genomic DNA fragments (Rauch et al., 2006). Serendipitously, this assay also provides indications on loci of MBD binding. The original analysis of this data revealed multiple changes in DNA methylation associated with ALL, with approximately two thirds of the differentially methylated regions (DMRs) being hypomethylated and preponderantly present at intergenic and intronic localization (Almamun et al., 2015). For the purpose of the current study, we focused on DMRs located at regions involved in alternative splicing.

In addition to extensive transcriptional reprogramming as described previously (Almamun et al., 2015), the ALL cells exhibited many variations of transcript isoform expression, since 2034 genes are differentially spliced with high confidence when comparing the patient cells to the HBC cells (Sup. Fig. S4C and Table S4). Almost 30% of the genes differentially spliced in ALL were also downregulated in their overall expression and 5% were upregulated. This suggests that in ALL, unlike what we observed in the DKO cells, a proportion of the alternative splicing could arise from transcriptional mis-regulation or result in non-sense mediated decay. Of note, we recovered the differentially spliced genes CD45/PTPRC and PKM which are dependent of DNA methylation and CTCF and CTCFL respectively (Shukla et al., 2011; Marina et al., 2016; Singh et al., 2017). DNMT genes were upregulated, while TET genes were downregulated ((Almamun et al., 2015) and Sup. Fig. S4E), indicating that any change in DNA methylation is likely not a simple consequence of variation levels in the products of these genes.
We next examined the 12 genes affected in their splicing by DNMT1 knock down in HeLa cells (Sup. Fig. S3K). Among these, 7 were expressed in ALL and pre-B cells, including 5, namely CD44, RABGAP1L, DST, GNAS, and GLS, displaying modified splicing between ALL patient cells and the HBCs. Among these, CD44 and GLS displayed changes in splicing involving the same exons and only CD44 showed variations consistent with those observed in the DNMT1-depleted HeLa cells.
Since CD44 appears as a potential marker linking cancer and DNA methylation, we analyzed its splicing in more details in ALL patients. As expression of CD44 was significantly reduced and more variable in the patient cells (Fig. 4A), we focused splicing analysis of this gene on the 9 ALL samples where the message was in the same range of the HCB controls (Sup. Fig. S4D). To evaluate the inclusion level of the CD44 variant exons in these RNA-seq samples, we used the same strategy than the one described for HCT116 cells. We observed that the number of reads covering junctions from variant to constitutive exons were decreased in ALL cells in comparison to normal HCB cells while reads spanning the C5-C16 junction of constitutive exons were significantly increased (Fig. 4B).
As in the HCT116 DKO cells and in the Hela cells with DNMT1 knock-down, the reduced inclusion of CD44 variant exons was associated with decreased levels of DNA methylation at several positions inside the body of the gene as illustrated by MIRA-seq data (Fig. 4C and Sup. Fig. S4F for individual tracks of MIRA-Seq). This data shows that CD44 variant exons may function as a sensor of intragenic 5mC variations occurring during tumor processes.

## Few alternative splicing events may function as sensors of DNA methylation

We next investigated whether DNA methylation would allow predicting alternative splicing events in addition to CD44. For that, we compared genes differentially spliced in ALL with those affected by DNMT1/3b inactivation in HCT116 cells (DKO). Among the differentially spliced genes in ALL, we identified 64 genes sharing at least one differentially regulated splice site with DKO cells. Manual curation with IGV and VOILA allowed identification of 49 genes subject to the same local variation of splicing in the two cellular models (Fig. 5A). The remaining genes included 10 cases of alternative promoters, and 5 cases where alternative splicing event are different in ALL and DKO cells even sharing a common splice site. Among the common splicing events, 75% (41/49) were located in the neighborhood of regions subject to differential methylation (DMRs). Among these, modified alternative splicing was consistent with the change in 5mC levels in ALL and in DKO cells only in 56% (23/41) of the cases, whereas 32% (13/41) were inconsistent with the variation of DNA methylation. This indicated that, statistically, DNA methylation was poorly predictive of the outcome of alternative splicing for the host genes under scrutiny.

Pathway analysis of the 23 genes for which a link existed between splicing and DNA methylation in both ALL and DKO cells revealed an enrichment in the KEGG pathway “spliceosome “ (ko03040, adjusted P value = 0.021) (Fig. 5B). In particular, these genes included the regulator of CD44 splicing TRA2B, and also the splicing factors TRA2A and PRPF38B, for which variations at a same DMR resulted in the same splicing event in both ALL and DKO cells (Fig. 5 D, E, F). The gene ontology (GO) analysis also identified enrichment in genes with a molecular function annotated as “GTPase activator activity” (GO:0005096, adjusted P=5.6E-3), including the genes ARFGAP2, ARHGAP12, RGS14 and APLP2 (Fig. 5C). As a control for these pathway analyses, examining the 26 genes from the remaining clusters (“opposite”/”5mC absent”/”uncertain DMRs”) did not yield enrichment in the pathways listed above (Sup. Fig. 5). This suggested that a subset of splicing factors or GTPase regulators could be directly sensitive to variations in DNA methylation through the modulation of their alternative splicing. Such sensors of DNA methylation may in turn trigger more global changes in alternative splicing associated with DNA methylation.

## Variant exon inclusion of CD44 is correlated with DNA methylation level in an in vitro model for breast-tumor progression.

The data we collected from the DKO cells, the HeLa cells depleted of DNMT1, and the ALL patient cells, all indicated direct and indirect effects of DNA methylation on pre-mRNA splicing. We noted that, in each model system, reduced DNA methylation inside the body of the CD44 gene translated into reduced inclusion of CD44 variant exons. However, increased inclusion of CD44 variant exons was mainly known as a marker of carcinoma metastasis and cancer stem cells (Orian-Rousseau, 2015). To determine whether this correlation was verified during epithelial cancer progression, we finally examined an *ex vivo* model of breast cancer featuring the non-tumorigenic human breast epithelial cells MCF10A cells, cells derived from MCF10A by transformation with activated RAS(T24), and two cell lines derived therefrom that reproducible form either Ductal Carcinoma In Situ (DCIS)-like lesions or metastatic carcinomas (CA1A) in xenografts (So et al., 2012). This series of cell lines, sharing a same genetic background and origin, constitute a good model for tumor progression to compare different cellular and molecular properties.

RT-qPCR assays showed that transition from MCF10A to the more aggressive downstream tumor cell populations resulted in only minor increases in the global expression of CD44 (Fig. 6A). In contrast, inclusion of variant CD44 exons was drastically increased (Fig. 6B, left panel), and symmetrically, levels of transcripts skipping all the variant exons were decreased (Fig. 6B, right panel). This modified processing of the CD44 pre-mRNA was initiated upon transformation of the MCF10A by RAS, then persisted during the subsequent steps of tumour progression. The overall pattern of DNA methylation in MCF10A, that was very similar to that observed in HeLa cells (compare Fig. 6C to Fig. 3C). Corroborating a positive correlation between the inclusion of variant exons and levels of DNA methylation, MeDIP experiments showed that in all cell lines derived from MCF10A, DNA methylation was increased at several locations inside the body of the CD44 gene (Fig. 6D). These changes in DNA methylation occurred mostly outside of highly methylated regions, the main contributors being variant exons and their intervening introns (Table 1). Thus, our results, obtained with four independent cellular models, indicated that intragenic DNA methylation appears predictive of the outcome of CD44 alternative splicing.

# *DISCUSSION*

In the present paper we wished to examine the relationship between deregulation of alternative splicing and the modified DNA methylation frequently observed during tumorigenesis. In the case of the CD44 gene, we observe a very general correlation between DNA methylation levels and inclusion of alternative exon in four different model systems. However, when examining the potential causes of this relationship, it was clear that we had to consider both direct effects locally mediated by chromatin, and indirect effects mediated by differential activity of transcription and splicing factors. Several mechanisms have been proposed to explain a direct connection between DNA methylation and alternative splicing; the best documented being an implication of methyl-binding proteins (MBDs).

## How DNA methylation can directly modulate the alternative splicing

In this context, two factors, namely CTCF and MeCP2, have been examined in depth (Shukla et al., 2011; Young et al., 2005; Maunakea et al., 2013; Wong et al., 2017; Zheng et al., 2017). In the case of CD44 alternative splicing, CTCF is unlikely to be involved since we did not observe binding sites nor recruitment in or near the CD44 variant region in either of the cell lines we examined, including HCT116, DKO (ChIP-seq Sup. Fig. S2B), and HeLa cells (ChIP-seq from ENCODE). In addition, we observe that reduced DNA methylation reduces inclusion of CD44 variant exons, while a roadblock mediated by CTCF recruitment after demethylation of the DNA would be expected to increase inclusion of these exons.
As for MeCP2, we find that depletion of this protein does not have a clear inhibitory effect on CD44 variant exons in HeLa cells (Fig. 2). This is apparently incoherent with earlier observations from MeCP2 mutant mice where CD44 is found aberrantly spliced in cerebral cortex mRNA (Young et al., 2005). But this difference may be dependent of neuronal-specific splicing factors absent in HeLa cells.
The role of other MBDs in splicing has not been yet characterized and we showed here for the first time that several MBD family members are at the interface between methylated DNA and splicing/chromatin factors. In particular, MBD1 depletion resulted in a profound inhibition of CD44 variant exon inclusion (Fig. 3E), while its recruitment to the CD44 gene body correlated with levels of DNA methylation (Fig. 2E). Likewise, depletion of MBD2 also had an important negative effect on CD44 variant exon level. Interestingly, this MBD is present on methylated exons/gene bodies of active genes and is thought involved in the modulation of RNAPII pausing (Menafra et al., 2014). Thus, reduced levels of MBD2 on the CD44 gene body is expected to alleviate potential RNAPII pausing, a process that would favor the skipping of CD44 variant exons that we observe (Batsché et al., 2006; Saint-André et al., 2011). MDB3 does not bind preferentially to methylated DNA (Saito and Ishikawa, 2002), but is recruited via MBD2 and is therefore found enriched on exons of active genes (Baubec et al., 2013; Gunther et al., 2013). MBD3 also co-purifies with in vivo assembled U2 snRNP spliceosomal complex in a mass spec analysis (Allemand et al., 2016). This suggests that MBD3 recruitment to gene bodies may in part be dependent on splicing decisions, alike what was shown for H3K36me3/SETD2.

## Relation with H3K9me3 and HP1γ

Several studies have suggested that MBD1-mediated transcriptional repression relies on the recruitment of HP1α and histone methylases, including SUV39H1 (Fujita et al., 2003) and SETDB1 (Ichimura et al., 2005). Remarkably, the distribution of H3K9me3 within CD44 in HCT116 cells (Fig. 2E) resembled closely that observed in HeLa cells (Saint-André et al., 2011; Ameyar-Zazoua et al., 2012), *i.e.* with the higher enrichment covering the end of the region containing the variant exons (v8, v9, and v10 ). This pattern is different from that observed for 5mC within CD44, as variant exon v9 and v10 are particularly poor in CpGs, and therefore in 5mC (Fig. 3C, 3D, 6C, Sup. Fig. S3C). This alone evidences the absence of positive correlation between H3K9 and DNA methylation. Earlier studies have suggested that H3K9me3 may compensate for 5mC in gene body regions with low CpG density (Hahn et al., 2011; Hon et al., 2012). A same mechanism of compensation may be at play in the body of genes for the regulation of alternative splicing.

The inverse correlation between levels of H3K9me3 and CpG density also argues against an implication of histone methylation marks in the guiding of DNA methylases to the coding region of CD44. However, we have found that the level of H3K9me3 within the endogenous chromatin of CD44 variable exons is decreased in absence of DNA methylation in DKO cells (Fig. 2E). This is in accordance with reintroduction assays of exogenous methylated or unmethylated DNA template in the chromatin that showed meDNA-dependency of H3K9me3 (Hashimshony et al., 2003; Yearim et al., 2015). We observed also that, even decreased, a part of the H3K9me3 level remains high which is also in accordance with the relative stability of the pericentromeric H3K9me3 levels in DNMTs null ES cells (Lehnertz et al., 2003). In fact, in CD44 we have earlier showed that H3K9me3 deposition at the end of the variant region is guided by an Argonaute-dependent chromatin associated complexes (Ameyar-Zazoua et al., 2012) and we note that in the DKO cells, expression of AGO1 is increased (Fig. 2B).

Even though little evidence exists for a connection between HP1γ recruitment and DNA methylation in the regulation of splicing (Yearim et al., 2015), this link may be more complex than anticipated. Indeed, in HCT116 cells, we observe that accumulation of HP1γ on the chromatin of the CD44 gene body is dependent on the level of DNA methylation (Fig. 2E), yet independent of the levels of H3K9 methylation, especially where the H3K9me3 levels is weak and unmodified in DKO. On most of the CD44 gene body, it seems that in the DKO cells, the loss of 5mC renders HP1γ accumulation independent of H3K9me3. Altogether, these data suggest that recruitment of HP1γ to the CD44 gene body which modulate the RNAPII kinetic (Saint-André et al., 2011), may be dependent of MBD1 (and possibly of MBD2/3) in CpG containing region, while it relies on H3K9me3 in regions with low CpG density (i.e. v9 and v10) (Fig. 7A). These data also suggest that DNA methylation is not required to cover exactly the sensitive variant exon since a certain degree of spreading of the chromatin readers is reached via regulation of RNAPII kinetic or protein-protein interactions.

## Discrimination between the direct and indirect effects of DNA methylation on the modulation of alternative splicing

We observed that of loss of DNA methylation in somatic human cells is associated with extensive changes in alternative splicing. Yet, the analysis of gene expression that we carried out in parallel suggested that the DNA methylation-splicing correlation may largely be rooted in a complex combination of indirect effects due to modified expression or splicing of splicing regulators (Fig. 2A) and transcriptional regulators (Table S2A).

Here, by documenting modified expression of ESRP1 and other factors upon loss of DNA methylation, we bring evidences for an implication of splicing factors and transcriptional regulators in the splicing modifications observed in the DKO cells. Thus, regulatory loops linking expression of splicing/chromatin factors to the overall chromatin state may also contribute to splicing deregulations imputed to local chromatin changes. It is noticeable that there is a very good correlation between the inclusion of CD44 variant exons and the levels of DNA methylation locally in three different cell-types : HCT116/DKO, DNMT1-depleted HeLa cells and in MCF10A-derived breast tumor cells.

As short-term depletion of DNMT1 in HeLa cells had little effect on the transcriptome, including no detecTable change in the expression of splicing factor, it offered good opportunity to explore the direct effect of DNA methylation on alternative splicing. Depletion of this DNA methylase modified splicing only at a limited number of genes.

Yet, several of these genes were highly relevant for cancer and EMT (Sup. Fig. S3K). These genes included CD44 and DST required for anchoring intermediate filaments to hemidesmosomes in epithelial cells and involved in human metastatic processes, the kinesin KIF1B (Gordon et al., 2019), the Glutaminase GLS (Lee et al., 2016b), and GNAS, a stimulatory G-protein alpha subunit (G s-α) mutated in colon cancer and associated with familial non-medullary thyroid cancer(Rochtus et al., 2016).

Most of the ALL-related changes of DNA methylation observed in gene bodies were a decrease as reported by (Almamun et al., 2015). This hypomethylation of genes bodies is thought to be a consequence of a passive dilution during uncontrolled cell division of tumor transformation (Wild and Flanagan, 2010). This has consequences not only on the pool of expressed genes but also on the quality of the mature transcripts since more than two thousand genes were found differentially spliced in these leukemias (Sup. Fig. S4D). DNMT1/3b depletion in HCT116 cells to some extend mimics this phenomenon and allows us to identify the few alternative splicing events that can be linked to a decrease of DNA methylation. Since these splicing events occur in very different cell-types with different developmental origins, they are likely to be a consequence of changes in DNA methylation. Differentially spliced genes detected in both DKO cells and in ALL within regions where DNA methylation was perturbed, showed a strong enrichment for genes taking part in RNA binding, GTPase regulatory networks, and cell adhesion processes (Fig. 5B). This indicates that the DNA methylation status of cells is linked to the activity of the RNA processing machinery, to GTPase regulators and to the cell-cell/cell-ECM interactions through splicing modifications. This is potentially important as it suggests that the genes subject to splicing perturbations in cancers are also the ones participating in cellular interactions within the tumor niche (Bastos et al., 2017; Fonseca et al., 2018)

## Genes subject to DNA methylation-dependent splicing in different cells share common molecular functions

An important conclusion from this study is that DNA methylation, that at best was considered as a way of fine-tuning alternative splicing, may for small number of genes, have an impact such that the level of DNA methylation present inside the body of these genes can have predictive value when estimating the splicing isoforms produced in a given cell type.
This is further supported by a recent study showing that splicing variations at the single cell level can be accurately predicted based on local DNA methylation (Linker et al., 2019).
Another intriguing possibility, and perhaps a more interesting one, is that these predicTable genes may serve as sensors of DNA methylation to adapt gene expression to external cues (Fig. 7B). In other words, some RNA binding proteins/splicing factors such as PRPF38B, TRA2B, TRA2A, by changing their exon composition, may translate changes in DNA methylation into modified activity of the splicing machinery. A similar mechanism may apply for other pathways/cellular functions such as metabolic enzymes (as GTPase) or membrane proteins (as CD44, CD47, APLP2, NISCH). It is also remarkable that TRA2A/B splicing factors have been described as binding factors of MeCP2 (Wong et al., 2017; Zheng et al., 2017). Further investigations will be required to know whether/how specific isoforms of TRA2A or TRA2B can bind differentially to MeCP2 for instance, and mediate particular alternative splicing decision. The question of the potential differential activity of TRA2b isoforms on the CD44 alternative splicing will also require further investigations.

In conclusion, we have discovered a causative role of DNA methylation in the regulation of alternative splicing via MBD1, MBD2 and MBD3 factors. Our findings reveal that most variations of alternative splicing are consequences of indirect effects. However, careful examination of splicing and intragenic methylation in two different cellular contexts also bring us to suggest that a limited number of splicing factors are directly affected by the intragenic DNA methylation and may function as mediators of some of the indirect effects. These splicing factors may also contribute to regulation of splicing of genes directly affected by intragenic 5mC, creating a potential positive loop of regulation between chromatin modifications and splicing factor expression. Altogether, we propose that the relationships between DNA methylation and alternative splicing decisions is an integration of multiple intricate and subtle mechanisms. In parallel, some genes including CD44 are directly sensitive to intragenic 5mC level. In this context, the variation of intragenic DNA methylation occuring during tumorigenesis, may be predictive of alternative splicing outcome of these genes.

## *MATERIAL AND METHODS*

### Cell culture, siRNA transfections, and PMA treatment

Colorectal carcinoma cell line HCT116 (CCL-247) and double knock-out (DKO) cells for DNMT1 and DNMT3b have been purchased from Horizon Discovery. HeLa cells (CCL-2) were from ATCC. MCF10A, MCF10A(ras), DCIS and CA1A were a kind gift from Annick Harel-Bellan (INSERM, France). Cells were maintained in Dulbecco's modified Eagle's medium (Gibco) supplemented with 7% (v/v) fetal bovine serum (Thermo Scientific) and 100 U/ml penicillin-streptomycin (Gibco). MCF10A were grown in the same medium supplemented with 0.5 µg/mL hydrocortisone, 10 µg/mL Insulin and 20 ng/mL EGF. HeLa cells were transfected with a mix containing siRNA (20 nM) and RNAi Max reagent (Life Technologies) according to the reverse transfection protocol then collected 72 h after for analysis. For experiments of DNMTs depletion, the cells were transfected twice, the second round of transfection was performed 48 h after the first one, and the cells were collected 72h after. siRNAs were from Qiagen or SIGMA, and their sequences are available upon request. PMA was purchased from SIGMA and resuspended in DMSO at 1 mM. Cells were treated with 10 nM of PMA.

### RNA extraction, reverse transcription and PCR:

Cytoplasmic enriched total RNA was extracted by homogenizing the cells in TMS buffer (10 mM TRIS pH8, 250 mM NaCl, 1 mM MgCl2 , 1% NP40, 20 mM DTT, 1U/µL RNasin) for 15 min on ice. After quickly spinning out the nuclei and debris, supernatant was supplemented with 0.5% SDS and 2 mM EDTA before extraction by phenol-chloroform, and isopropanol-precipitation. RNase-free TURBO DNase (Ambion; Sigma) -treated RNAs were checked on agarose gel for their quality. Reverse transcription was performed using M-MLV and affinityScript (Agilent) reverse transcriptases with oligo dT and random hexamers. PCR products of semi-quantitative polymerase chain reaction (PCR) were resolved on agarose gel and verified by DNA sequencing. Quantitative real-time PCR (qPCR) was assayed in 10 µl reactions with Brillant III Ultra Fast SYBR-Green Mix (Agilent) using a Stratagene MX3005p system. The analysis was performed using the MxPro software. The sequences of primers used for PCR are available upon request.

### Affymetrix exon array:

Transfected HeLa cells with siRNA targeting DNMT1 or GAPDH in independent biological triplicate were extracted. RNAs were hybridized on GeneChip Human Exon 1.0 ST Arrays (Affymetrix), and scanned following the manufacturer's instructions. Analysis was performed by GenoSplice (http://www.genosplice.com). Only genes expressed in the cells for at least in one of the two compared conditions and giving a signal of good quality probes were considered for further analysis. Significant variations (P< 0.05) in exon variation above 20% were taken into account. Suspected events were counted manually in Arrays data visualized on the EASANA® visualization interface

### Antibodies :

Antibodies were purchased from Diagenode for 5mC antibody (3D33), MBD1 N-terminal (#078-050), MeCP2 (#052-050), H3K27ac (C15410174), from Active motif for DNMT1 (#39-906), MBD2 (#39-548), from Abcam for RNAPII pS2 (ab5095), histone H3 (Ab1791), H3K9me3

(ab8898), from Euromedex for HP1a (2HP-2G9-AS), BRG1 (2SN-2E12-AS), from Millipore for HP1γ (42s2, 05-690), H3K9me2 (07–441), from Santa-Cruz for Sam68 (7-1, sc-1238), ASF/SF2 (sc-33652), U2AF65 (MC3, sc-53942), RNAPII (N20, sc-899), N-cadherin (CDH2, sc-7939), from Cell Signaling for SUV39H1 (D11B6, #8729), from SIGMA for EHMT2 (HPA050550), from ThermoFisher for ESRP1 (PA5-25833), E-cadherin (CDH1, PA5-32178), Vimentin (PA5-27231), CD44v5 (VFF-8, MA5-16967), and from eBioscience for CD44v7-v8 (VFF-17, BMS118). Hermes3 ascite directed against CD44 constant exons was a kind gift from Larry Sherman and Peter Herrlich (University of Karlsruhe, Germany)

## Protein extraction, cell-fractionation, western blot and immunofluorescence.

Protein was prepared by boiling cells 10 min in lysis buffer containing 50 mM TRIS pH8, 150 mM NaCl, 0.5 mM EDTA, 1% SDS and protease cocktail inhibitors (Roche). Proteins were separated by electrophoresis on 4–12% gradient PAGE gels and transferred on nitrocellulose membrane for western blot. Cell fractionation of proteins resolved on western blot and immunofluorescence were performed as previously described (Mauger et al., 2015).

## Chromatin (ChIP) and Methylated DNA immunoprecipitation (MeDIP)

ChIP assays were performed as previously described (Ameyar-Zazoua et al., 2012; Mauger et al., 2015). For MeDIP assay, nuclei from RNA extracted cells in TMS buffer were resuspended in 300µL TRIS pH8 10 mM, EDTA 2 mM, SDS 0.5% and proteinase K (0.8 mg/mL), incubated 5h at 55°C before purification by phenol (pH7)/chloroform and precipitation by isopropanol and NaAcetate at room temperature during 10 min. Genomic DNA was recovered by low centrifuge speed 10 min / 4700 rpm, treated by RNase A (20 µg/mL) for 1h at 37°C in TE buffer and sonicated 8 min with BioRuptor (Diagenode) (15 sec ON / 15 sec OFF) at low intensity and at 4°C. Sheared DNA was checked on agarose gel electrophoresis to be around 500 bp. DNA were boiled for 10 min, chilled 10 min on ice, and diluted in IP buffer (10 mM Na-Phosphate(pH 7.0), 140 mM NaCl and 0.05 % Triton X-100). One µg of DNA was incubated for 4h with 1 µg of 5mC antibody 3D33 or non-immune IgG, followed by a 2 h incubation with 40 µL of anti-Mouse-magnetic beads (Dynabeads). Beads were washed 3 times for 5 min with 1 mL of IP buffer, and 2 times in TE NP40 0.01%. Beads were eluted by boiling 10 min in 100 µL H2O containing 10% (V/W) chelex resin (BioRad), PK-digested for 30 min and then finally incubated 10 min at 95°C. Equivalent to 0.5 µL was used for qPCR assays.

## Bioinformatic analysis

6 RNA-seq libraries from wild type and DKO (double knockouts *DNMT1 DNMT3b)* HCT116 cells were obtained from 4 different studies (Simmer et al., 2012; Schrijver et al., 2013; Maunakea et al., 2013; Blattler et al., 2014) (listed in Sup. Fig S1C). RNA-seq libraries from acute lymphoblastic leukemia (ALL, n=19) and pre-B healthy controls (n=8) were as well obtained from the SRA database (Almamun et al., 2015). Reads quality and statistics were assessed using FASTQC 0.10.1 (https://www.bioinformatics.babraham.ac.uk/ projects/ fastqc/), MultiQC 1.0 (Ewels et al., 2016) and in house python script. After a first alignment using STAR 2.5.0a (Dobin et al., 2013), principal component analysis of gene expression based on the 500 most variable genes evaluated by DESeq2 with Rlog normalization indicated two separable clusters for ALL samples (Sup. Fig. S4A). The cluster of 7 ALL RNA-seq (SRR2031977, SRR2032039, SRR2032043, SRR2032105, SRR2032108, SRR2032111, SRR2032112) contained low quality libraries with a percentage of uniquely mapped reads

<60% and a percentage of reads assigned to an annotation <10% (Sup. Fig. S4B). Consequently, these data were not included in our further analyses.

Genome index was produced with STAR using Ensembl GRC37 release 75 primary genome assembly and annotations. According to STAR manual and for more sensitive novel junction discovery, the junctions detected in a first round of mapping were used in a second mapping round. To discard reads potentially originating from pseudogenes, reads were mapped with only one mismatch allowed and multi-mapper reads were not counted.

When not clearly stated in the original study, read strandedness was inferred using infer_experiment.py from the RseQC 2.6.4 package (Wang et al., 2012). ALL dataset was further considered as un-stranded. Read counts were then computed with featureCounts 1.5.2 (Liao et al., 2014) at the gene meta-feature level. The obtained counts were further investigated for evidences of gene differential regulation with DeSeq2 1.14.1 (Love et al., 2014). Principal component analysis is based on logCPM counts corrected with TMM method and obtained from EdgeR 3.16.5 (Robinson et al., 2010). Analysis for HCT116 and DKO cells was conducted as explained (Sup. Fig. S1D). Alternative splicing analysis was conducted using Majiq 1.0.4 (Vaquero-Garcia et al., 2016) using the annotation provided with the software and the mappings obtained with STAR separately for ALL, polyA DKO and totalRNA DKO datasets and with polyA and totalRNA DKO datasets grouped together (Sup. Fig. S1F).

38 MIRA-seq alignments from the Almamun et al. study were downloaded from SRA (Almamun et al., 2015). Broad DNA methylation peaks were detected using MACS 2.1.1 (Zhang et al., 2008). Differentially methylated regions of interests (ROI, ie peaks detected by macs2) between ALL and healthy samples were identified using R bioconductor MEDIPS 1.24 package (Lienhard et al., 2014) using the EdgeR method which calculates scale factors using the TMM method. Bigwigs were generated from Almamun et al. (2015) alignments using deeptools (Ramírez et al., 2016) . Raw coverage from the MIRA-seq alignments were described graphically with in-house python scripts.

## *ACKNOWLEDGEMENTS*

The authors thank Annick Harel-Bellan for kindly provide valuable reagents. We are grateful to our colleagues C. Rachez and E. Allemand for helpful discussions, Leslie Landemaine, and Leslie Dupont for helpful preliminary experiments, Catherine Bodin for technical assistance and Edith Ollivier for administrative assistance.

## *FUNDING*

Ministry of Research (MENRT; O.M.); ERASMUS program and University (B. H. and C. H.-L.) The Centre National de Recherche Scientifique (CNRS; E.B. and C.M.). Agence Nationale de la Recherche (ANR-11-BSV8–0013); REVIVE—Investissement d'Avenir (to C.M., O.M. and E. K.). Funding for open access charge: CNRS recurrent funding.

Conflict of interest statement. None declared.

**Figure 1: Loss of DNA methylation leads to Epithelial-to-Mesenchymal Transition (EMT) and a decrease of variant CD44 protein isoforms**.

**A**) HCT116 cells WT or DKO cells were fixed, permeabilized and subjected to immunofluorescence with the indicated mouse antibodies against CD44 (Hermes3) recognizing an epitope in the C5 exon (Goldstein et al., 1989), v5-CD44 (VVF8), v10-CD44 (VVF14) and v7v8-CD44 (VVF17), and with rabbit antibodies specific to E (CDH1) or N-(CDH2) cadherins. Anti-CD44 mouse antibodies are revealed in red, and anti-cadherin rabbit antibodies are in green. Blue staining is DAPI. The right panel shows a western blot analysis of E-cadherin (CDH1) and Vimentin (VIM) expression.

**B, G**) RNA-seq from 4 studies were combined (n=5 samples) and matched for studies and types of extracts. The average of duplicate from the Blatter's study has been considered as one sample, the s.e.m. is indicated in the error bar (see SupFig S1C). The relative levels were cpm normalized for the libraries sizes. Statistical analysis has been proceeded on Rlog (DESeq2) normalized counts using a paired t-test (two-tail) to compare WT to DKO levels. The indicated genes in the top panel are mesenchymal-specific genes and in the bottom panels are epithelial-specific genes.

**C**) Map of the human CD44 gene and mRNA.

**D**) RT-PCR of total extract from HCT116 WT or DKO treated (+) or not (-) by PMA for 6 h. The primers and sequenced products of each PCR assay are indicated with their size (bp) on the right. PCR for RNA with constitutive exons are shown as control.

**E)** Relative mRNA levels of CD44 variant exons in HCT116 WT and DKO cells. Total RNA from WT and DKO cells were extracted at each indicated day after plating. Equal amounts of RNA were treated by DNase, reverse transcribed and subject to RT-qPCR using specific primers of the indicated genes spanning at least 2 different exons. Relative quantities were normalized between WT and DKO for each individual day using non-modified genes (RPLP0, CDK9, CCNT1 or SAM68). Data are averages (error bar are s.e.m.) of 2 individual experiments.

**F**) Whole protein extracts from HCT116 WT or DKO treated (+) or not (-) by PMA for 20 h, were analysed by western blot using antibodies against constant exons of CD44 (Hermes3) in the top panel and against v5 exon of CD44 (VVF8) in the middle panel. The bottom panel shows the ponceau-coloured membrane as loading control. A 1/10 diluted extract of MCF10A was used to indicate the size of variant CD44 isoforms (CD44v). An arrow highlights the constitutive CD44s isoforms and the main isoforms bringing the v5 exons.

**G**) The inclusion of CD44 variant exons, from the RNA-Seq studies. Differential splicing was evaluated by counting the reads covering the indicated exon-exon junctions.

The relative levels of each junction are indicated as the percentage of inclusion or skipping junctions among the total junctions involving the C5 exon (top panels) or involving the C16 exon (bottom panels). Significance were evaluated by using Wilcoxon signed-rank test, for $\alpha=0.05$ (two-tails). # indicates that there is sufficient evidence to suggest a difference between DKO and WT cells. The differences were also evaluated using a Student's t-test for matched-samples (two-tails), with -$p<0.05$(*), $p<0.01$(**).

.

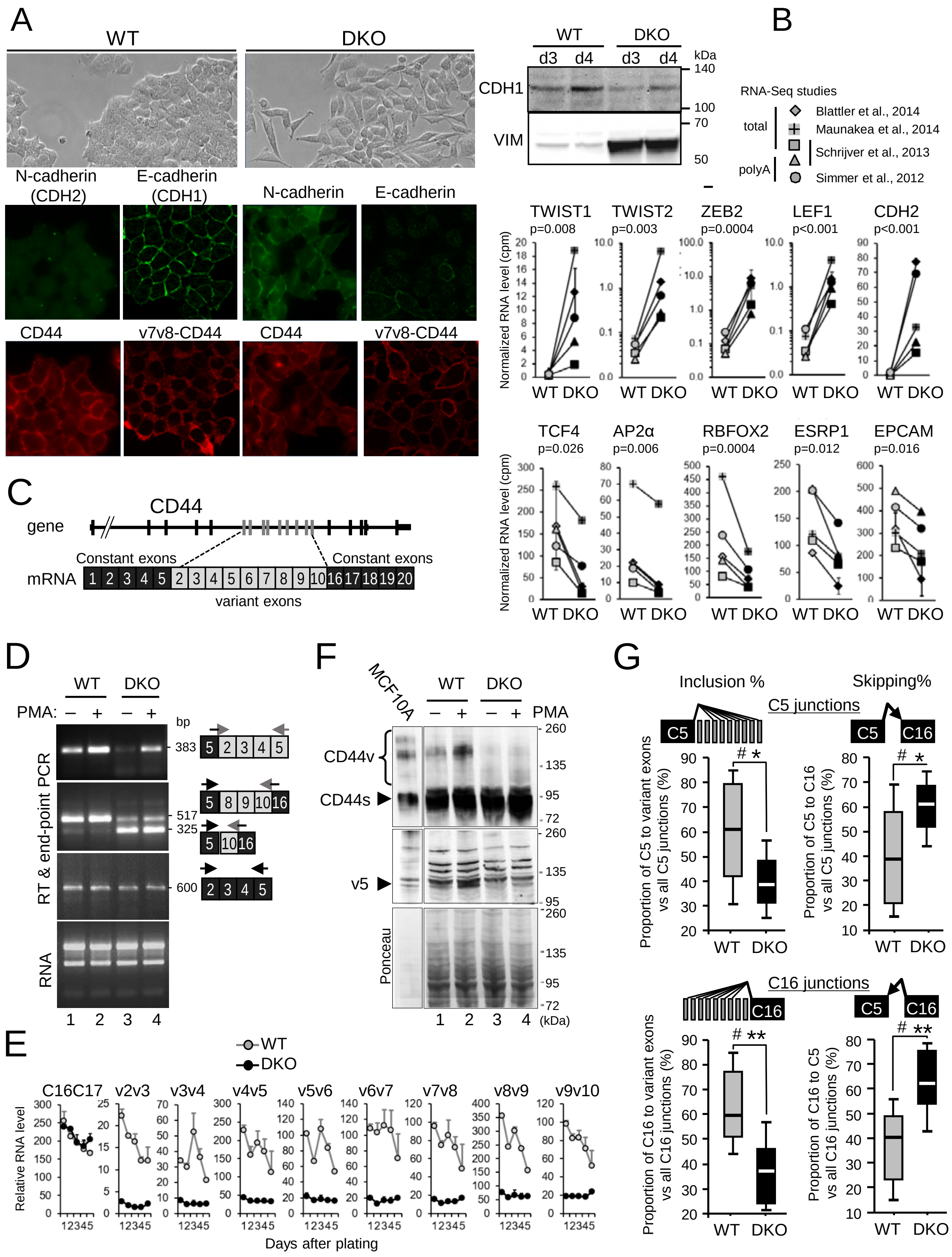

Batsché *et al.*, Figure 1

**Figure 2 : Modified expression and splicing of regulators of transcription and splicing in DKO HCT116 cells**.

**A**) Number of genes found differentially expressed or spliced in DKO versus WT HCT116 cells in the meta-analysis of 4 independent studies containing 6 RNA-seq datasets for each cell. Genes were considered differentially expressed with a log2 fold change >1 and p-value <0.05 (paired Test) as explained in Sup Fig S1C, S1D. The MAJIQ package (Vaquero-Garcia et al., 2016) was used to detect differentially spliced genes with high confidence P(|dPSI|>0.2)>0.95, (differential Percent of Splicing Index) between the DKO cells and the parental HCT116 cells, as explained in Sup Fig S1F. The list of the tested RNA binding proteins is available in Table S1

**B**) Transcriptional levels of indicated genes evaluated by RT-qPCR in DKO (black line) and WT (grey line) HCT116 cells harvested over 5 days after plating. Relative levels were expressed as described in Fig 1F.

**C**) Whole protein extracts from WT or DKO HCT116 cells, 3 and 4 days after plating were analysed by western blot using antibodies directed against the indicated proteins.

**D**) Skipping of the EHMT2 variant exon 10 in DKO cells. Top panel shows semi-quantitative RT-PCR performed on total RNA extracts and already described primers detecting the inclusion of e10 or its skipping (Δe10) (Mauger et al., 2015). Middle panel shows a western blot of EHMT2 isoforms in total protein extract. Bottom panel shows the ponceau-stained membrane as loading control. Ratio of EHMT2 protein spliced isoforms in cells were quantified by Image J using two independent sets of extracts on non-saturated revelations.

**E**) ChIP walking experiments were carried out with chromatin from DKO (black bars) and WT (grey bars) HCT116 cells, with antibodies for H3, H3K9me3, H3K9me2, HP1γ , and MBD1. Immunoprecipitated DNA was quantified by qPCR targeting indicated regions. Relative amount of H3K9me3 and H3K9me2 are expressed in percent of H3 and amounts of HP1γ and MBD1 are expressed relatively to the signal obtained for ChIP using non-immune IgG. Values are means ± s.e.m. of at least three independent experiments. Statistical significance of the differential levels were evaluated using Student's t-test (two-tailed), with p<0.05(*), p<0.01(**), p<1E-3(***). D, distal promoter region.

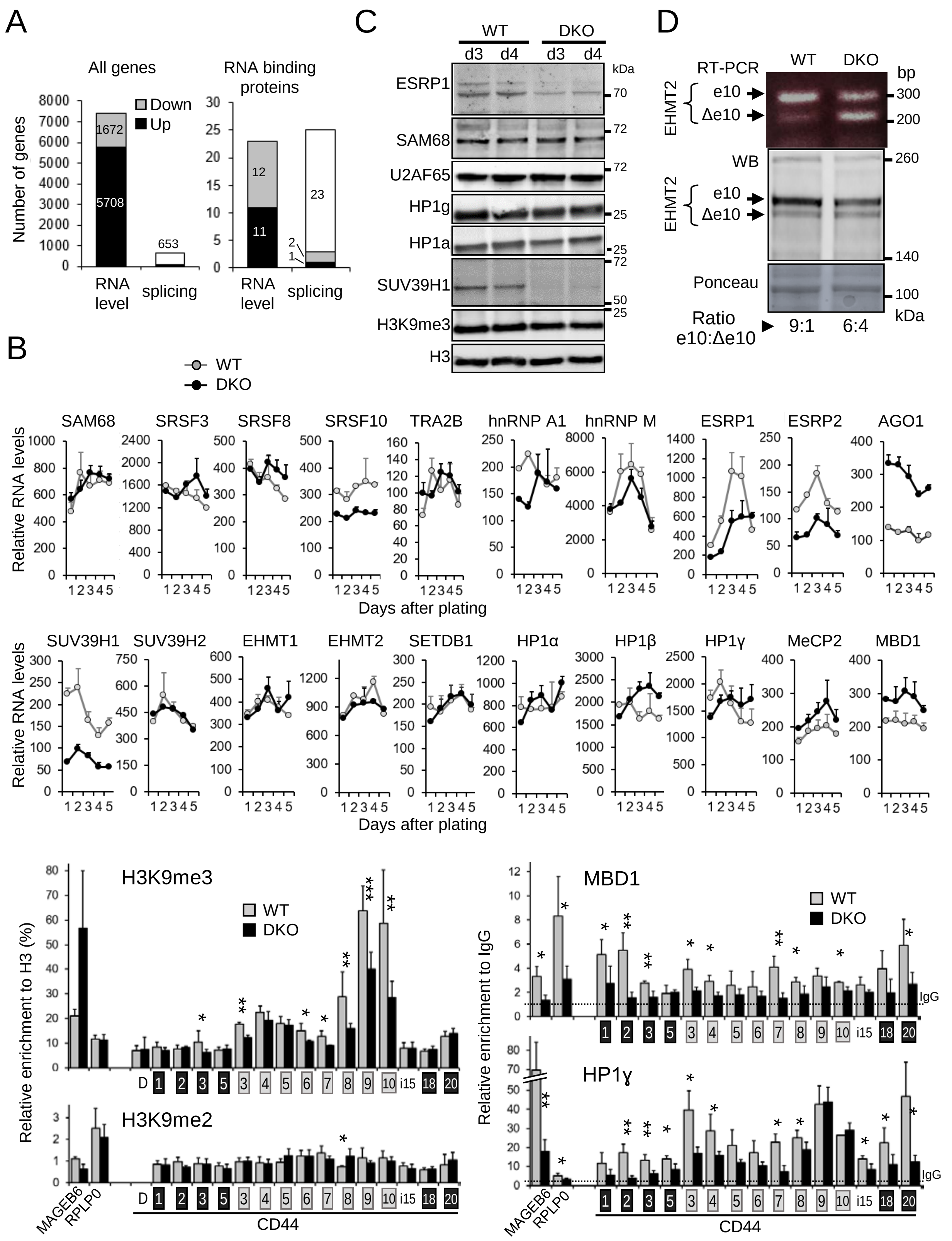

Batsché *et al.*, Figure 2

**Figure 3: CD44 variant exon level is dependent of DNA methylation in DNMT1 depleted HeLa cells**.
**A, B**) HeLa cells were transfected with two individual siRNAs or a commercially available pool of 3 siRNAs that target DNMT1, and treated with or without PMA (2h).
**A**) Protein extracts were subject to western blot analysis using the indicated antibodies. (see SupFig S1B for PMA-treated cell extract)
**B**) RNA extracts were subjected to RT-qPCR with primers amplifying the indicated exons. Relative RNA levels normalized with the RPLP0 reference gene, were used to calculate the fold change relative to the mock untreated cells (set to 1, dashed line). Statistical significance of the exonic differential levels upon DNMT1 depletion (without PMA) were evaluated using Student's test (two-tailed), with p<0.05(*), p<0.01(**), p<1E-3(***). The right panel shows the PMA effect (blue bars) in these conditions of transfection with the pool of DNMT1 siRNAs. There was no significative difference between siRNA NT transfected cells and the mock cells when only considering constitutive exons. The variant exons were increased by PMA in presence of DNMT1.
**C, D**) CD44 intragenic DNA methylation is dependent of DNMT1 expression and independent of PKC activation.
**C**) DNA from HeLa cells transfected with the DNMT1 pool of siRNAs were analyzed by methylated DNA immunoprecipitation (MeDIP) using the 3D33 antibodies directed against methylated DNA (5mC) or non-immune IgG as negative control. The levels of 5mC was expressed as a percentage of the input DNA quantities for each of the qPCR primers covering the CD44 gene : intronic sequences (i) are indicated by white boxes, constant exons are indicated by black boxes and variant exons by light grey boxes, D is a distal sequence upstream to the TSS. The germline-specific H2BWT gene was used as a positive control for DNA methylation in the somatic HeLa cells, while the GAPDH promoter allowed estimating the background signal yielded by a non-methylated CpG-rich region. The control IgG was at least 100 times less enriched and is not represented. The data were averages (+s.e.m) of two independent experiments.
**D**) HeLa cells treated with or without PMA for the indicated time and the relative enrichment of 5mC were evaluated by MeDIP as previously described.

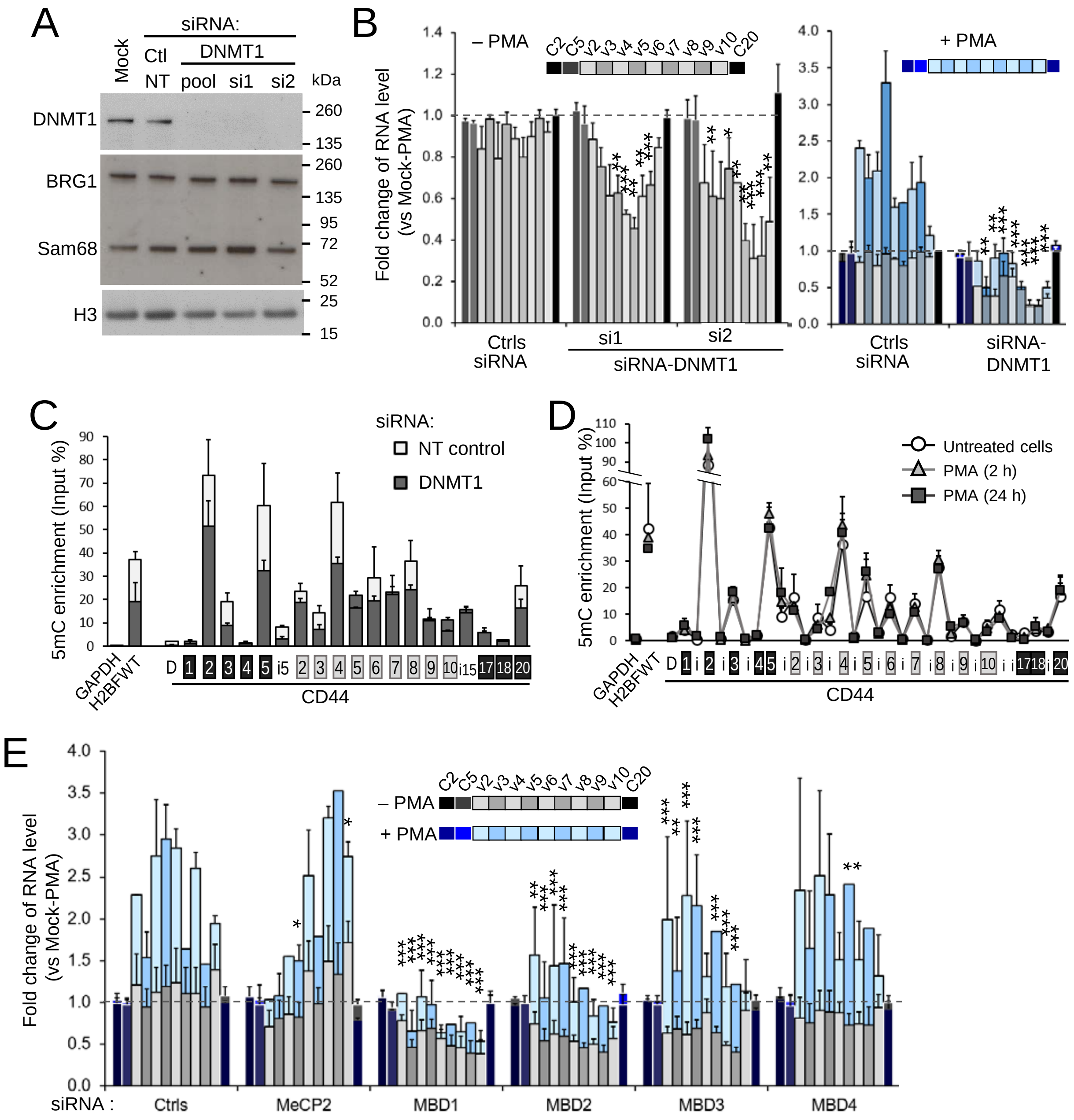

Batsché *et al.*, Figure 3

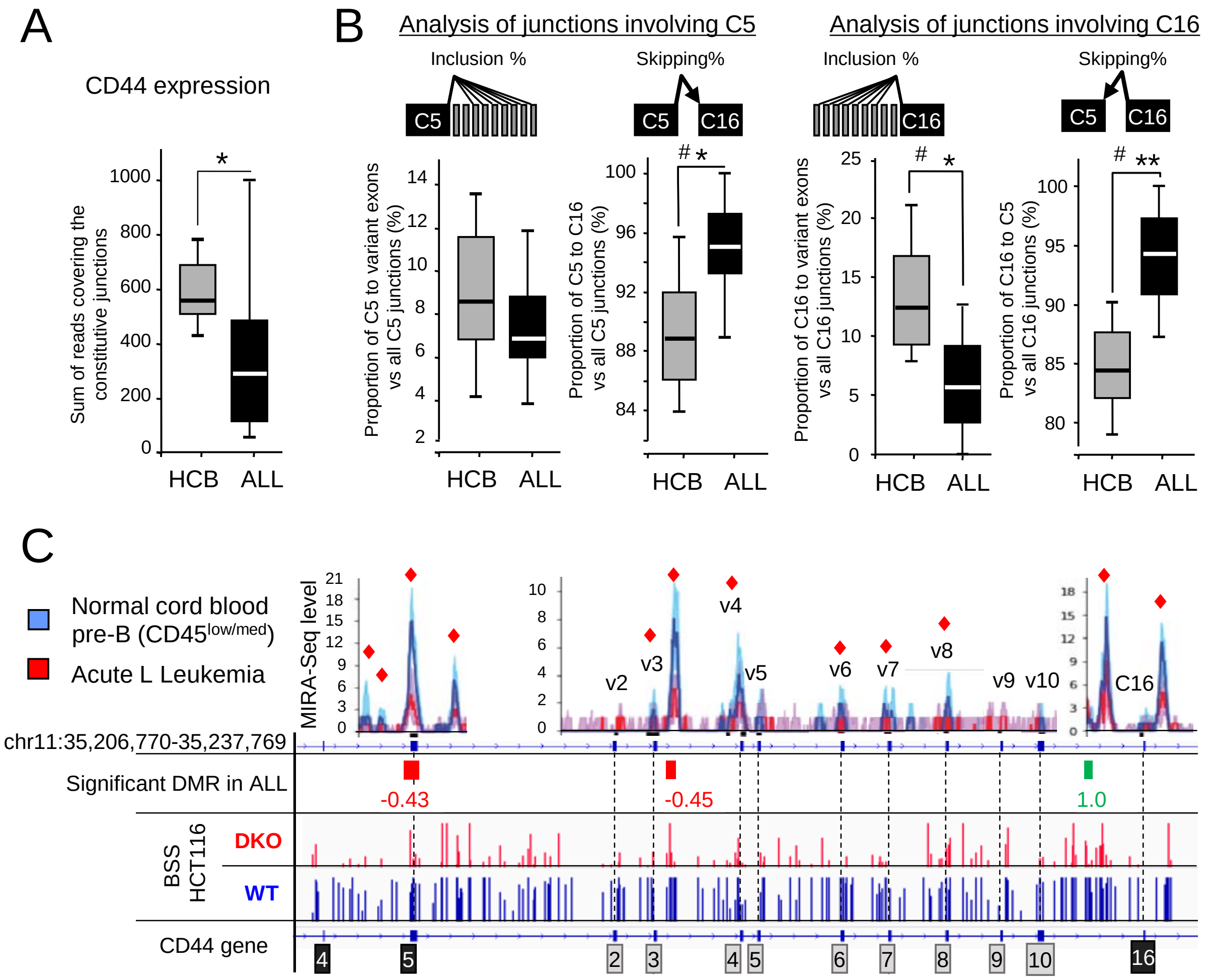

**Figure 4 : CD44 variant exons were significantly less included in ALL samples than in pre-B cells from Human Cord Blood (HCB).**

**A**) The CD44 global expression is evaluated by the normalized count sum of reads covering the constitutive exon-exon junctions for each of the 8 HCB control or the 9 selected ALL samples showing comparable levels of CD44 expression (Sup. fig. S4D).

**B**) The proportion of skipped and inclusion junctions were calculated by summing the counts of all the junctions, detected by at least two reads, between the indicated constant exon and all the variant exons. Significance was evaluated by using Student's t test (one-tail), where p-values are indicated as < 0.05(*) or < 0.01(**)). Significance was also evaluated using the Wilcoxon ranked test: where # indicates that there is sufficient evidence to suggest a difference between ALL and HCB cells with α=0.05 (one-tail).

**C**) The top graphs represented the bin levels of the MIRA-Seq counts from the 18 ALL samples(red) and the 20 pre-B samples (blue). Individual tracks are shown in Sup Fig S4F. The lines indicate the median while the shadowed areas represent quartiles. The red scares indicate significant difference between the level of bins with p<0.05. The log2 fold change of the Differentially Methylated Regions (DMRs) in ALL versus normal pre-B after normalization and benferroni correction for multitesting p<0.05. For comparison the RBBS from HCT116 and DKO cells are shown at the bottom.

Batsché *et al.*, Figure 4

**Figure 5 : A few differentially spliced genes may be an indication of DNA methylation and its biological relevance.**

**A**) Differentially spliced genes from DKO (table S2) and from ALL (table S4) were compared to identify common splicing sites. RNA events were categorized as indicated using visual with IGV of ALL MIRA-seq and with VOILA of the RNA events from DKO and ALL.

**B**) Pathway analysis was performed using the Enrichr tool (Kuleshov et al., 2016) and the 23 genes with splice changes were correlated with changes in DNA methylation (green quarter).

**C, D, E, F**). Examples of genes with common splicing events in DKO vs WT HCT116 cells and in ALL versus HCB pre-B cells. The overlaid MIRA-seq tracks set at the same scale, for ALL and HCB samples spanning the region containing the regulated cassette exon (in grey boxes). Changes in the DNA methylation levels were evaluated, and displayed as decreased (red boxes) or increased (green boxes) DMRs below. When found to be significant (pval<0.05) after normalization and Benferroni's correction for multitesting, log2 fold change is indicated and boxes are coloured. In addition, empty boxes indicates obvious DMRs. Below, the Methyl-seq (RRBS) from DKO and WT HCT116 cells is displayed, followed by the RefSeq map of genes. The graphs are the quantification of the RNA alternative splicing of each indicated variant exons detected with high confidence P(|dPSI|>0.2)>0.95 by MAJIQ. For each sample, counts of reads covering the indicated junctions were used to calculate the proportion of each junction involving the source exons (s) or the target exons (t), and represented as a percentage of inclusion (red and pink) and skipping (blue). The 6 RNA-seq samples (APLP2), or the polyA+ (TRA2A, TRA2B, PRPF38B) of DKO and HCT116 cells were used to calculate the averages (±s.e.m.) and the Student's t test (one-tail): $P < 0.05$ (*), $P < 0.01$(**), $P < 0.001$ (***). The 8 HCB samples and the 12 ALL samples selected in SupFig S4A and S4B, were used for the same calculation.

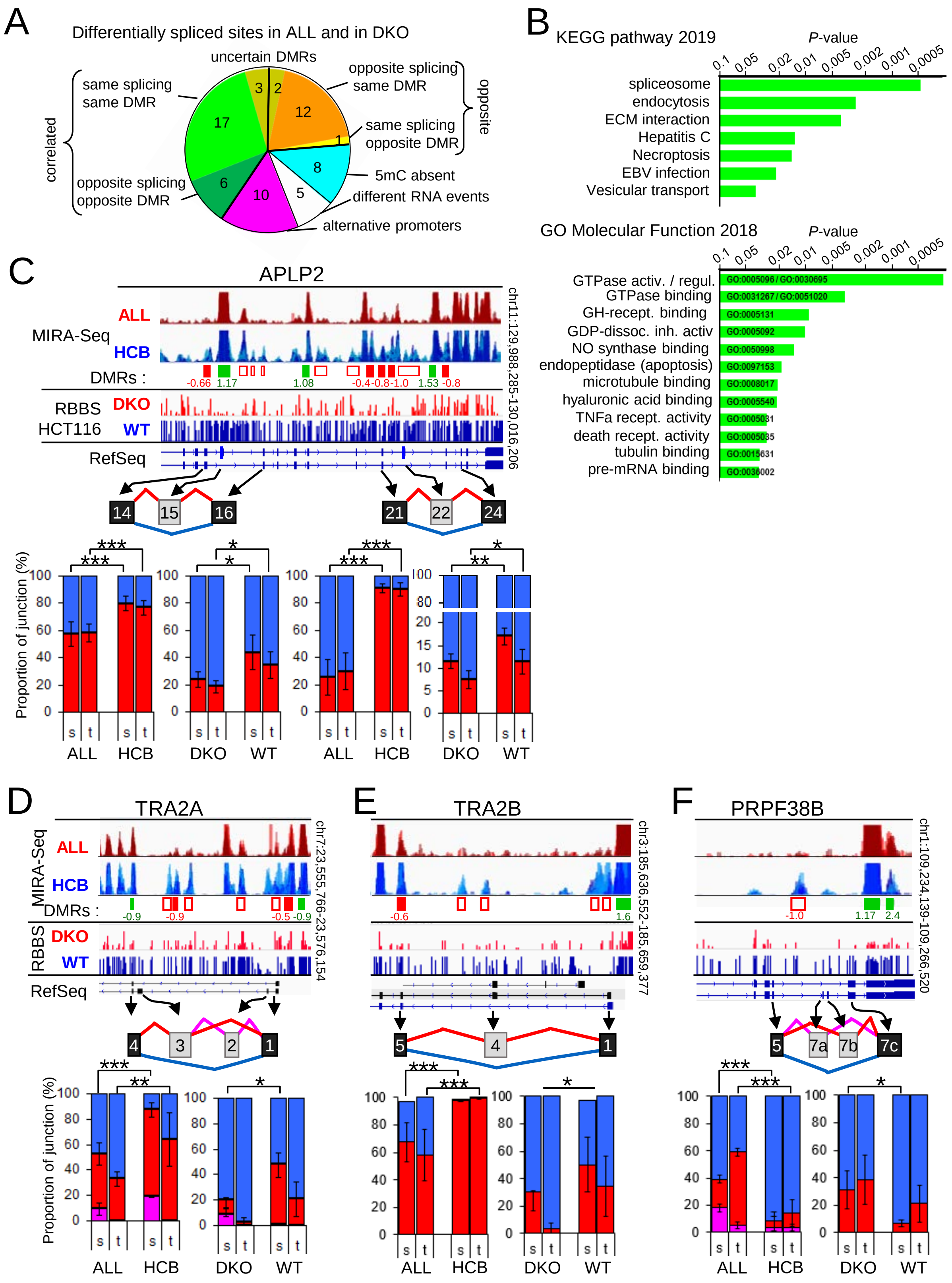

Batsché *et al.*, Figure 5

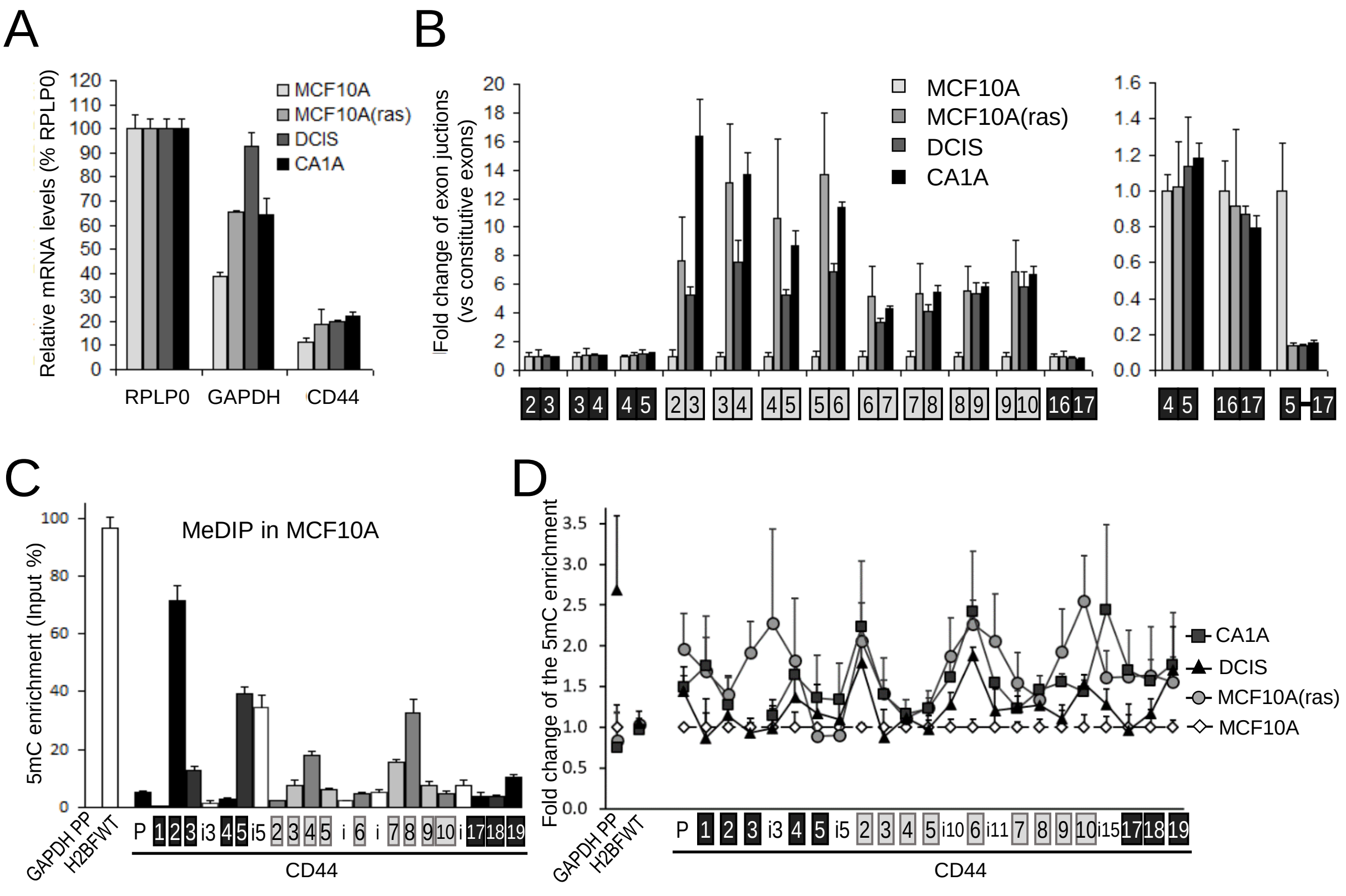

**Figure 6 : DNA methylation increases with the inclusion of CD44 variant exons in tumorigenic-derived MCF10A-cell lines**.

**A, B** ) RNAs extracted from cell lines were subjected to RT-qPCR using primers specific to the indicated regions. **A**) The CD44 mRNA level were the average of constitutive exons relative to RPLP0, which is used as an unmodified reference gene.

**B**) The levels of the variant exons (grey boxes) were normalised to the average of constitutive exons (black) and expressed as a fold change over the MCF10A level (set to one for each primer pair). The right panel indicates the fold change of the CD44 isoform skipping the variant exons (meaning the C5 to C16C17 exon junction).

**C,D**) MeDIP assays have been proceeded on purified DNA from the same cell-lines using the 3D33 antibodies directed against methylated DNA or non-immune IgG as negative control. Relative enrichment was corrected by the DNA amount of each input. The levels of DNA methylation in MCF10A parental cells is shown in **C**. The control IgG was at least 100 times less enriched and is not represented. **D**) The relative enrichment of meDNA in other MCF10A-transformed cell lines were expressed over the MCF10A level for indicated regions. The data were averages (±s.e.m) of two independent experiments with triplicates used for qPCR. The statistical analysis is in Table 1.

| Table 1 | Comparison vs MCF10A | | | vs DCIS |
|---|---|---|---|---|
| Differences on : | MCF10A(Ras) | DCIS | CA1A | CA1A |
| all locus tested | 1.4E-07 | 2.8E-07 | 5.7E-04 | 4.3E-05 |
| constant exons | 0.001 | 0.107 | 0.127 | 0.298 |
| variant region inluding introns (i10, i11, i15) | 1.1E-04 | 5.0E-04 | 0.006 | 0.004 |
| variant exons only | 0.003 | 0.005 | 0.029 | 0.012 |
| introns only | 0.035 | 0.050 | 0.043 | 0.068 |

**Table 1 : Statistical analysis of the differential levels of DNA methylation within the CD44 gene through the MCF10A-derived cells presented in Fig 6D**. Indicated Groups of loci for comparison were tested by pairwise Student's t test (two-tailed) evaluating differences from locus to locus. The p-values of these tests for each cell compared to MCF10A parental cells or CA1A versus DCIS are reported in the table. The non-significant p-values are in grey.

Batsché *et al.*, Figure 6

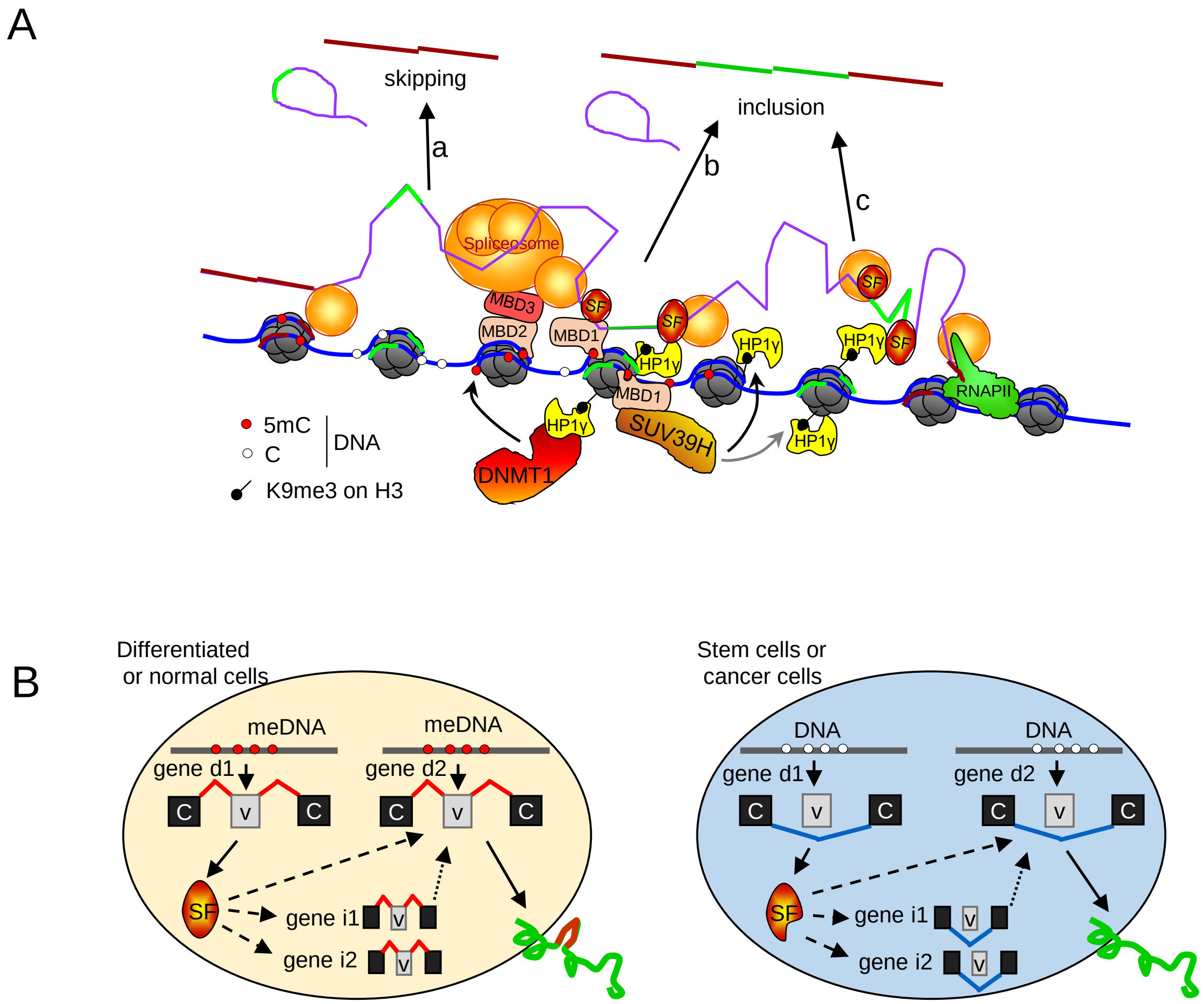

**Figure 7 : Model for the roles of the DNA methylation in the modulation of the alternative splicing.**
**A**) The absence of DNMT1/3b decreases the methylation of CpG (white circle), inhibits MBD recruitment, prevents the HP1γ spreading on the gene body, and favors the skipping of variant exons (a). DNA methylation (5mC) can be recognized by MBD proteins that may help regulate RNA alternative splicing by interacting with the spliceosome and/or splicing factors, (SF) or by promoting HP1γ recruitment on chromatin via H3K9me3 provided by SUV39H1 (b). In the absence of CpG, the high level of H3K9me3 still favours the recruitment of HP1γ locally (c) without counteracting the effect on splicing of the 5mC defect.
**B**) As directly impacted gene (gene d), some splicing factors, such as TRA2Aa/b, might play a role of a DNA methylation sensor by expressing specific isoforms that promotes alternative splicing decision on genes indirectly regulated (genes i) and also on genes directly regulated (genes d) by the DNA methylation. This last makes a positive loop reinforcing the modulation of alternative splicing by the local 5 mC levels.

Batsché *et al.*, Figure 7

# • Suplementary data

**<u>Supplemental Figure S1 : Double knock-out of DNMT1 and DNMT3b in HCT116 cells promotes epithelial-to-mesenchym transition (EMT).</u>**

**A**) Dot blot of methylated DNA from HCT116 WT or DKO using 3D33 antibodies

**B**) RT-qPCR of RNA from HCT116 WT or DKO showing reference genes in these cell series

**C**) Four studies-based meta-analysis of gene differential expression between HCT116 WT and HCT116 DKO (knock-out of DNMT1 and DNMT3b). Fastq files of indicated GEO data were aligned on the hg19 genome using the STAR software. Multimapper reads or reads with more than one mismatch were not considered, Software detection indicates that data were not oriented except for Schrijver2012(Schrijver et al., 2013) . The data from Maunakea et al. (Maunakea et al., 2013) were considered as not oriented even though software is unable to determine it. The number of reads uniquely aligned on hg19 are indicated in brackets for each dataset.

**D**) Test of normalization methods. After normalization the DKO and WT samples were tested for significant differences by using a paired T test (two-tailed) on the 5 comparisons as depicted in the table in **C**). The average of duplicate from the Blatter's study has been considered as one sample, the s.e.m. is indicated in the error bar in **E**). The limit of change has been considered significant when the mean of the fold change is >1.5 and the p-value of the paired-T-test on the log2(cpm) is <0.05. The log2 normalization allowed to detect a more important fraction of upregulated genes compared to the « Rlog » normalization (proceeded with DESeq2) used as reference. The union of the two normalization methods predicted more differentially expressed genes than the individual tests.

**E**) Normalized RNA level (count per millions of reads in librairies) of reference genes (DNMTs and H19). Expression of Vimentin (VIM), Desmoplakin (DSP) two markers of EMT and CD44 were shown to complete the **Figure 1**. RPLP0 is shown as a non-modified gene. The statistical paired T-test (two tail) was used to calculate the p-value on the log2 normalized counts and indicated when <0.05 (n.s., non significant).

**F**) Differentially spliced genes in RNA-seq of DKO cells compared to the parental HCT116 cells. In order to take into account the statistical dispersion of the data due to the differences of RNA extraction methods, three different comparisons as indicated have been conducted using MAJIQ algorithm. "Local Splicing Variation" (LSV) estimated by MAJIQ were used to detect differentially spliced genes with high confidence between the conditions P(|dPSI|>0.2)>0.95, (Vaquero-Garcia et al., 2016) where dPSI is the "differential Percent of Splicing Index". The number of predicted genes having at least one alternative splicing event is indicated for each comparison. Finally, all the genes differentially spliced with high confidence in each comparison were considered.

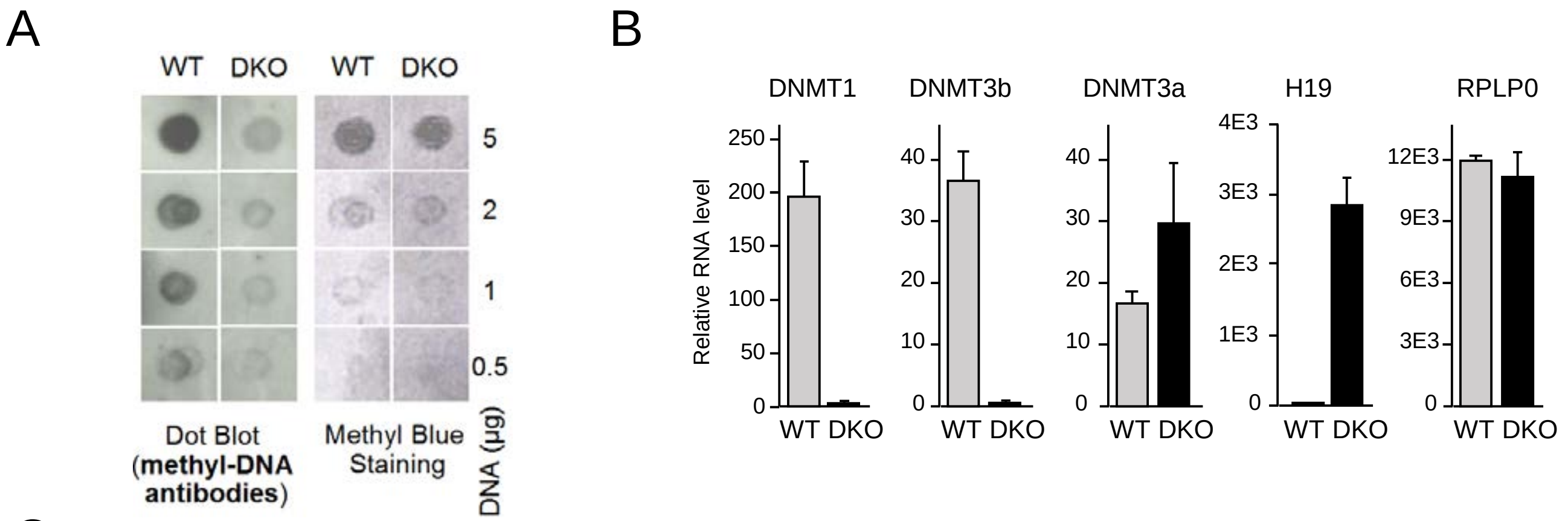

C

| RNA-Seq from : | RNA extract : | WT | DKO |
|---|---|---|---|
| Simmer et al., 2012 GSE39068 | polyA+ ● | SRR517830 (19 095 262) | SRR517831 (20 651 768) |
| Schrijver et al., 2013 GSE45332 | polyA+ ▲ | SRR787303 (30 609 125) | SRR787305 (29 480 692) |
| | total ■ | SRR787296 (13 879 253) | SRR787304 (15 741 044) |
| Maunakea et al., 2013 GSE47488 | total ✚ | SRR869304 (41 087 539) | SRR869305 (43 391 672) |
| Blattler et al., 2014 GSE60106 | total ◆ | SRR1030462 (36 281 078) | SRR1536578 (34 864 087) |
| | | SRR1030463 (37 769 873) | SRR1536577 (33 287 125) |

average

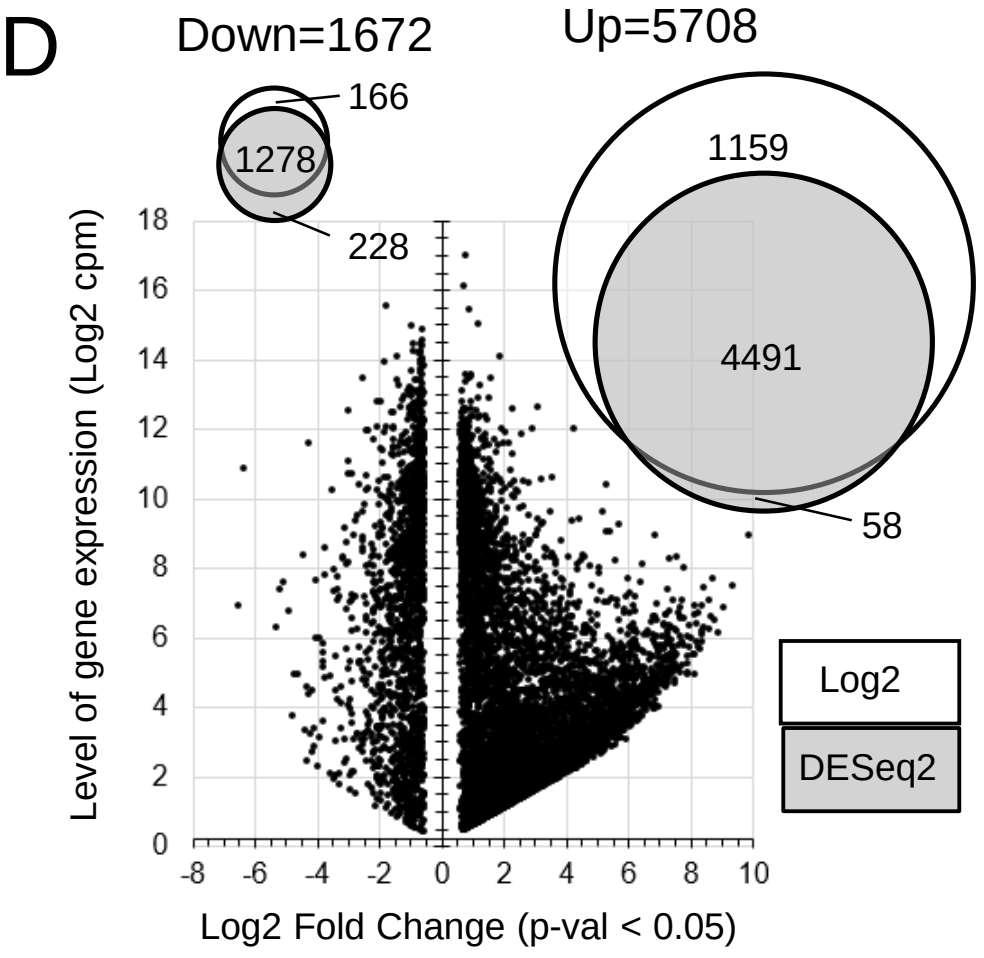

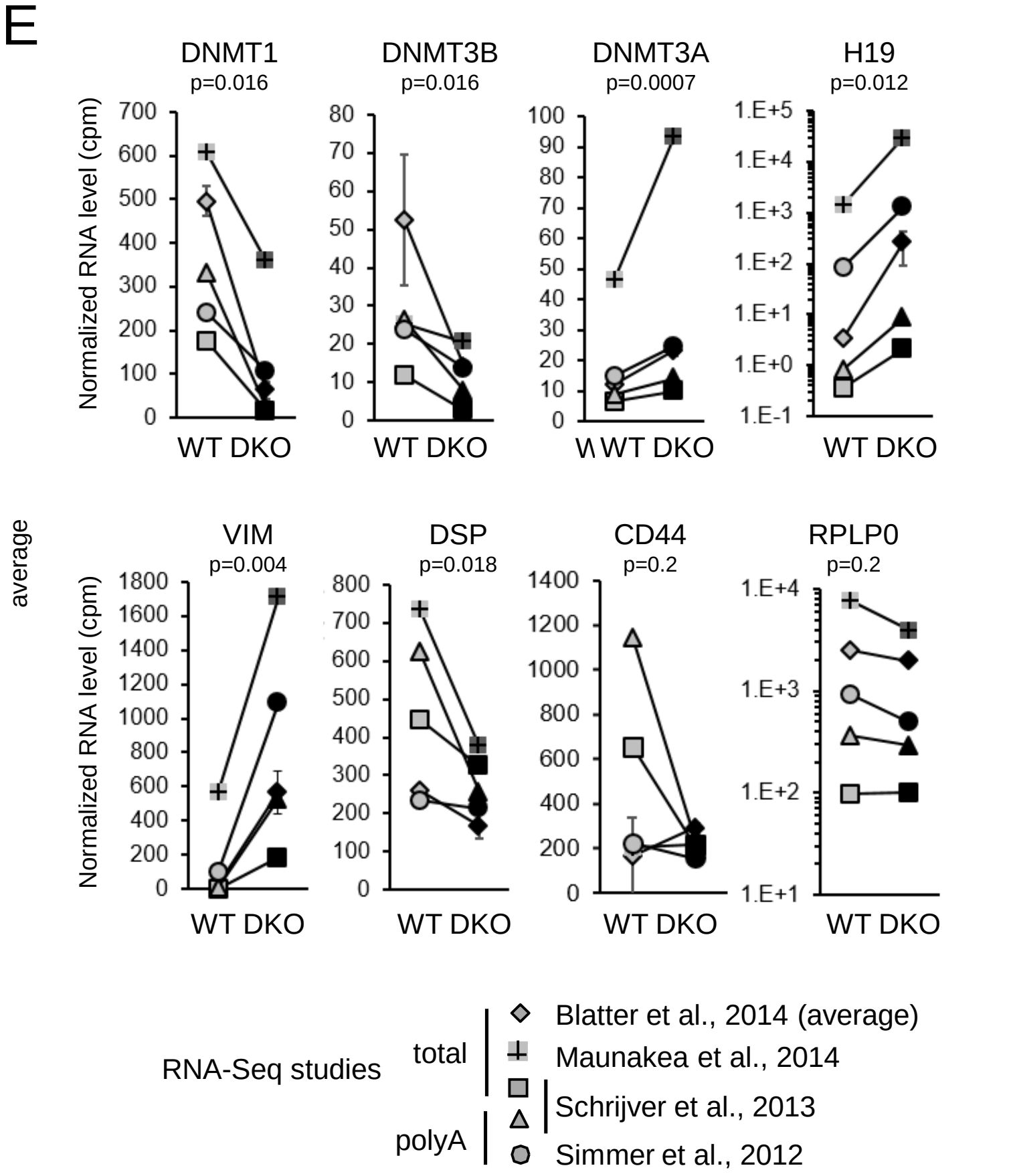

F

| RNA extracts | DKO vs WT | Genes |
|---|---|---|
| Total RNA | 4 vs 4 | 387 |
| polyA RNA | 2 vs 2 | 298 |
| All RNA | 6 vs 6 | 217 |
| Combinaison of all comparisons : | | 653 |

Batsché *et al.,* Suplementary figure S1

**<u>Supplemental Figure S2 : Meta-analysis of gene expression in HCT116 DKO cells, and histone modifications covering the CD44 gene.</u>**

**A**) Normalized RNA level (count per millions of reads in libraries) of indicated genes confirming the RT-qPCR of the **Figure 2** and completing the **Sup. Fig. S1E**. The statistical paired T-test (two tail) was used to calculate the p-value on the log2 normalized counts which are indicated only when <0.05 and the average of the fold change was >1.5.

**B**) IGV view of DNA methylation from MOMe-Seq (MOMe)and Bisulfite-seq (BSSEeq), and of ChIP-seq for the indicated histone marks (Lay et al., 2015) and CTCF (Maurano et al., 2015) covering CD44 in DKO and in WT HCT116 cells. Tracks were set at the same range for DKO and WT and are showed with HCT116 tracks below the DKO tracks for each indicated feature. Duplicated ChIP-Seq were overlaid on one track. The used datasets are listed here:

.

| | **WT** | | **DKO** | |
|---|---|---|---|---|
| BigWig files used from GSE58638 (files around 2 Go) (Lay et al., 2015) | | | | |
| H3K27Ac | | GSM1415876 | GSM1420161 | GSM1415884 |
| H3K4me1 | GSM1420154 | GSM1415875 | GSM1420160 | GSM1415883 |
| H3K4me3 | GSM1420153 | GSM1415874 | GSM1420159 | GSM1415882 |
| H2AZ | GSM1420152 | GSM1415873 | GSM1420158 | GSM1415881 |
| H3K36me3 | GSM1420157 | GSM1415879 | | GSM1415887 |
| MOMe-Seq HCG (250 Mo) | GSM1420150 | | GSM1420151 | |
| | | | | |
| BigWig files used from GSE58695 (files around 250 Mo) | | | | |
| Bisulfite-Seq HCG | GSM1416976 | | GSM1416977 | |
| BigWig files used from GSE50610 (files around 45 Mo) (Maurano et al., 2015) | | | | |
| CTCF | GSM1224649 | GSM1224650 | GSM1224654 | GSM1224655 |

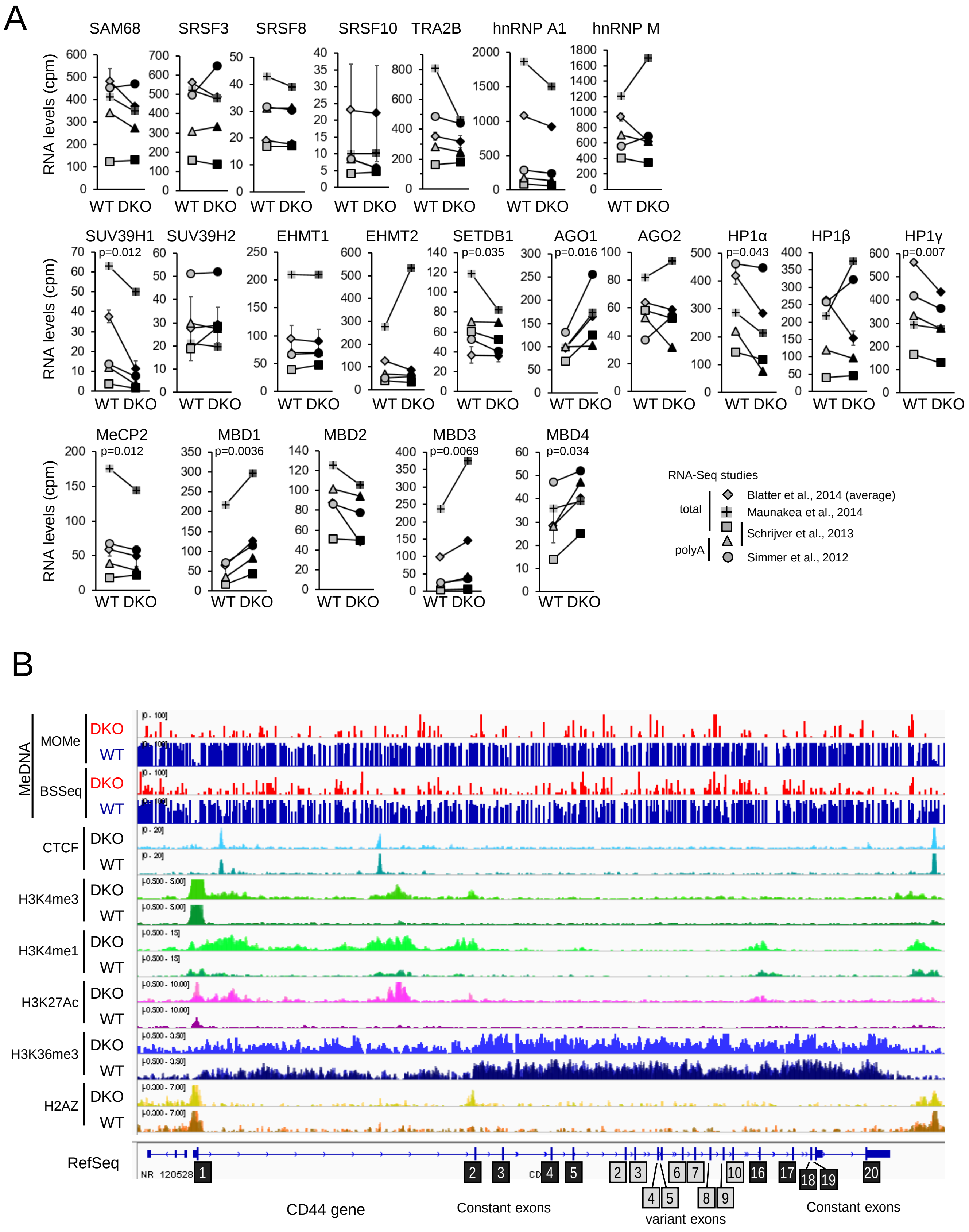

Batsché *et al.*, Suplementary Figure S2

# Batsché *et al.*, Table S2B

Differential expression and splicing of RNA-binding protein/splicing factors in DKO cells versus WT HCT116 cells. This table is a hub of Table S1 and Table S2A for splicing factors.

Regulated RNA binding factors

| Ensembl ID | Gene | Parent of pseudogene | RNA level Log2 cpm | Differential RNA expression: Regulation | Log2 FC | pval | Differentially spliced curation | LSV ID | (MAJIQ) LSV Type | Regulation dPSI>20% [95%] <-More in DKO // More in WT-> |
|---|---|---|---|---|---|---|---|---|---|---|
| ENSG00000109536 | FRG1 | p | 4.8 | | | | CE8 | 190878553-190878657:S | | 22 / 22 |
| ENSG00000107625 | DDX50 | p | 5.6 | | | | CE3 | 70666467-70666763:T | | 23 / 23 |
| ENSG00000126653 | NSRP1 | p | 6.2 | | | | CE9… | 28499560-28499616:T<br>:28443664-28443881:S | | 40 / 47<br>31 / 37 |
| ENSG00000116001 | TIA1 | | 6.2 | | | | 5'SS E10/Ee11 | 70443536-70443631:S | | 24 / 24 |
| ENSG00000013441 | CLK1 | | 6.4 | | | | FE1 or FE6 | 201724403-201724469:S | | 30 / 30 |
| ENSG00000198563 | DDX39B | | 6.6 | | | | 3'SS E6 | 31508099-31508441:S<br>31509727-31510225:T | | 35 / 32<br>33 / 33 |
| ENSG00000117360 | PRPF3 | p | 6.6 | | | | CE4 | 150300234-150300925:T | | 20 / 20 |
| ENSG00000100296 | THOC5 | | 6.7 | | | | CE2+3 & CE15 | 29924926-29925228:S<br>29949660-29950243:T | | 30 / 30<br>39 / 39 |
| ENSG00000164548 | TRA2A | p | 7.4 | | | | CE4 | :23571408-23571660:T | | 30 / 36 |
| ENSG00000134186 | PRPF38B | | 7.5 | | | | CE7 | 109240322-109241449:T<br>109238899-109238959:S | | 13 / 23<br>22 / 22 |
| ENSG00000179950 | PUF60 | | 7.7 | | | | 3'SS CE6 | 144911450-144912029:T | | 13 / 22 |
| ENSG00000197111 | PCBP2 | p | 7.7 | | | | 5'SS FE | 53835525-53835584:S | | 26 / 26 |
| ENSG00000153914 | SREK1 | | 7.7 | | | | CE6 or TE | 65449396-65449618:S | | 27 / 27 |
| ENSG00000151923 | TIAL1 | | 7.9 | | | | 3'SS e7 | 121347664-121347760:T<br>121336288-121336417:T | | 21 / 25<br>41 / 41 |
| ENSG00000029363 | BCLAF1 | p | 8.1 | | | | CE13+14 | 136590575-136591097:T | | 30 / 28 |
| ENSG00000196504 | PRPF40A | | 8.2 | | | | CE11 | 153533965-153533989:S<br>153535643-153535986:T | | 20 / 20<br>20 / 20 |
| ENSG00000154473 | BUB3 | p | 8.3 | | | | TE8 or TE9 | 124922128-124922757:S | | 37 / 37 |
| ENSG00000145833 | DDX46 | | 8.3 | | | | 5'SS e22 | 134152120-134152296:S | | 32 / 32 |
| ENSG00000136527 | TRA2B | | 8.4 | | | | FE1 or FE2 | :185644389-185646861:S<br>:185655613-185655924:T | | 32 / 32<br>29 / 25 |
| ENSG00000124193 | SRSF6 | p | 8.6 | | | | CE7 or TE | 42089343-42092245:T | | 13 11 06 |
| ENSG00000135829 | DHX9 | p | 8.6 | | | | CE4 | 182821368-182821479:T<br>182811680-182811812:S | | 35 / 1710<br>38 / 38 |
| ENSG00000160710 | ADAR | | 8.9 | | | | CE4 | 154574861-154575102:S | | 40 / 42 |
| ENSG00000168566 | SNRNP48 | | 5.6 | down | -0.7 | 0.035 | CE5a | 7599906-7601757:T | | 17 / 26 |
| ENSG00000060138 | YBX3 | p | 7.7 | down | -0.9 | 0.006 | CE12+13 | 10856622-10857037:S | | 20 / 20 |
| ENSG00000092847 | AGO1 | | 6.9 | up | 0.7 | 0.016 | FE1 or FE2 | 36354028-36354211:T | | 77 / 77 |
| ENSG00000100320 | RBFOX2 | | 6.8 | down | -1.3 | 0.000 | | | | |
| ENSG00000152601 | MBNL1 | | 7.0 | down | -1.3 | 0.018 | | | | |
| ENSG00000136231 | IGF2BP3 | p | 5.8 | down | -1.2 | 0.011 | | | | |
| ENSG00000099622 | CIRBP | | 7.8 | down | -1.1 | 0.004 | | | | |
| ENSG00000104413 | ESRP1 | | 6.6 | down | -1.0 | 0.013 | | | | |
| ENSG00000065978 | YBX1 | p | 7.5 | down | -0.9 | 0.010 | | | | |
| ENSG00000148690 | FRA10AC1 | | 5.3 | down | -0.9 | 0.002 | | | | |
| ENSG00000137944 | CCBL2 | | 4.5 | down | -0.8 | 0.001 | | | | |
| ENSG00000056097 | ZFR | p | 7.1 | down | -0.8 | 0.010 | | | | |
| ENSG00000117614 | SYF2 | p | 6.0 | down | -0.6 | 0.006 | | | | |
| ENSG00000100056 | DGCR14 | | 4.4 | up | 0.6 | 0.038 | | | | |
| ENSG00000123136 | DDX39A | p | 7.2 | up | 0.6 | 0.038 | | | | |
| ENSG00000169217 | CD2BP2 | p | 6.1 | up | 0.7 | 0.011 | | | | |
| ENSG00000131043 | AAR2 | | 5.1 | up | 0.7 | 0.037 | | | | |
| ENSG00000071859 | FAM50A | | 6.2 | up | 0.7 | 0.043 | | | | |
| ENSG00000126803 | HSPA2 | | 2.6 | up | 1.2 | 0.003 | | | | |
| ENSG00000204389 | HSPA1A | | 2.4 | up | 3.8 | 0.010 | | | | |
| ENSG00000154548 | SRSF12 | | -0.7 | up | 4.1 | 0.002 | | | | |
| ENSG00000185272 | RBM11 | | -1.2 | up | 5.4 | 0.002 | | | | |
| ENSG00000128739 | SNRPN | p | -1.6 | up | 5.4 | 0.001 | | | | |

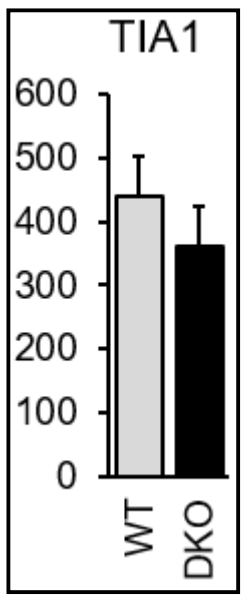

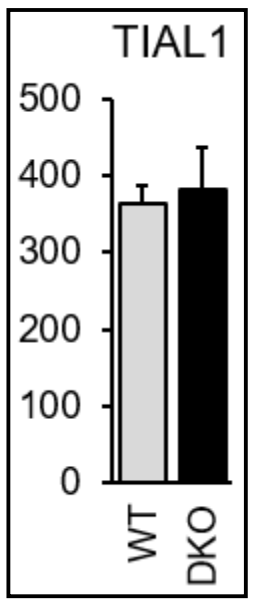

In the Manual curation column : CE=cassette exon, FE=First exon (promoter), TE=terminal exon (3'end), SS= splice site
In the LSV ID column: T=target, S=source
P indicated that these genes have been described as parent of pseudogene (Pei et al., 2012 in Genome Biol. 13,9,R51, The GENCODE pseudogene resource), which means that the splicing analysis can be misled by the expression of the retropseudogene or pseudogene.
Inset shows RT-qPCR validation of RNA-Seq analysis

**Supplemental Figure S3 : DNA methylation effects on alternative splcing in HeLa cells**
**A, B, D**) Knock-down efficiencies of DNMTs in HeLa cells. Transfection of HeLa with the indicated siRNA targeting DNMT1 were used to extract total RNAs or proteins. **A, D**) RNAs were extracted and subjected to RT-qPCR to quantify the RNA levels of DNMT genes. Relative RNA levels were expressed as fold change over the mock control. Non-targeting (NT) and GAPDH siRNA are used as negative controls. The right panel shows the relative levels of RNA corresponding to the imprinted H19 gene expressed as percent of RPLP0 reference gene. The two bars correspond to different primer pairs amplifying two separated regions of the transcripts.
**B**) Prior to extraction, cells were treated with PMA for 2h and DNMT1 and histone H3 proteins were revealed by western blot.
**C**) The proportion of CpG in the indicated loci ±200 bp surrounding the displayed qPCR amplicons in the MeDIP assays. CpG were counted on both DNA strands and expressed as percent of total dinucleotides.
**E**) Relative levels of CD44 variant exons are decreased upon depletion of DNMT1 but not of DNMT3A or DNMT3B. Levels of each indicated exons were normalized by the average levels between mock and NT siRNA of corresponding exons. Data are average (± s.e.m.) of at least three independent experiments. Statistical comparison have been were evaluated using Student's t-test (two-tailed), with $p<0.05$(*), $p<0.01$(**) , p<1E-3(***).
**F**) RNA levels of indicated genes in HeLa cells. RT-qPCR relative levels are expressed in percent of RPLP0.
**G**) Cell fractionation of the HeLa cells prepared as described in (Mauger et al., 2015). Equivalent volume of extracts were loaded on a SDS-PAGE and proteins were revealed on the membrane using the indicated antibodies.
**H, I**) Knock-down efficiency of indicated siRNA in HeLa cells. Three days after transfection, cells were extracted and the RNA levels were evaluated by RT-qPCR (**H**). Relative levels are expressed relative to the mock-transfected transfected cells (set to 1). Ctrl siRNAs are the mean of non-targeting and GAPDH. The data are averages (+s.e.m.) of at least four independent experiments. **I**) MBD1 protein level were evaluated by Western blot. The ponceau-stained membrane is shown as loading control.
**J**) Transcriptome-wide analysis of DNMT1 depleted HeLa cells by Affymetrix exon arrays. In three independent assays, transfected HeLa cells by the indicated siRNAs targeting DNMT1 and by GAPDH siRNA used as negative controls during 5 days have been extracted. Total RNA extracts were used on exon arrays. Gene expression levels have been calculated by the means of the triplicate on all the exons after normalization (analysis conducted by GenoSplice). The differentially spliced genes were detected by calculating for each exon a splicing index (the exon level versus the gene level) and comparing them in DNMT siRNAs versus GAPDH siRNA. The genes were considered differentially expressed or spliced with a fold change >1.5 and p-value <0.05.
**K**) List of the differentially spliced genes found with the two different DNMT1 siRNAs. Genosplice analysis of exon arrays calculated average signal of the 4 probes (at least 4) for each exon on the triplicates. These levels were compared to the average signal of corresponding genes to evaluate their Splicing Index (PSI). Exon changes were considered significant with a fold change>1.2 and P-value <0.05 for each individual siRNA. Visual curation using Genosplice interface allowed selecting RNA events common for the two siRNAs.
**L**) Exon array covering CD44, GLS and DST, according GenoSplice visualization. Each bar corresponds to one probe, and four probes were used for each exon; Bar height (log 2 scale) corresponds to signal intensity for the indicated condition and colors correspond to the change ratio of the GAPDH siRNA control: up-regulated in red, down-regulated in green, and unmodified in black. The changes of RNA after PMA-treatment of the cells were shown as positive control of variant exon increasing for CD44. Background signal is indicated by the yellow line. Exon numbers are indicated below, in grey rectangles. Maps of GLS and DST genes are showed indicating the annotated alternative splicing event (red lines), alternative promoters (red arrows) and alternative termination (noted "pA"). 3'UTR and 5'UTR are symbolized by red boxes in the maps and the thinner grey boxes below the probes.

.

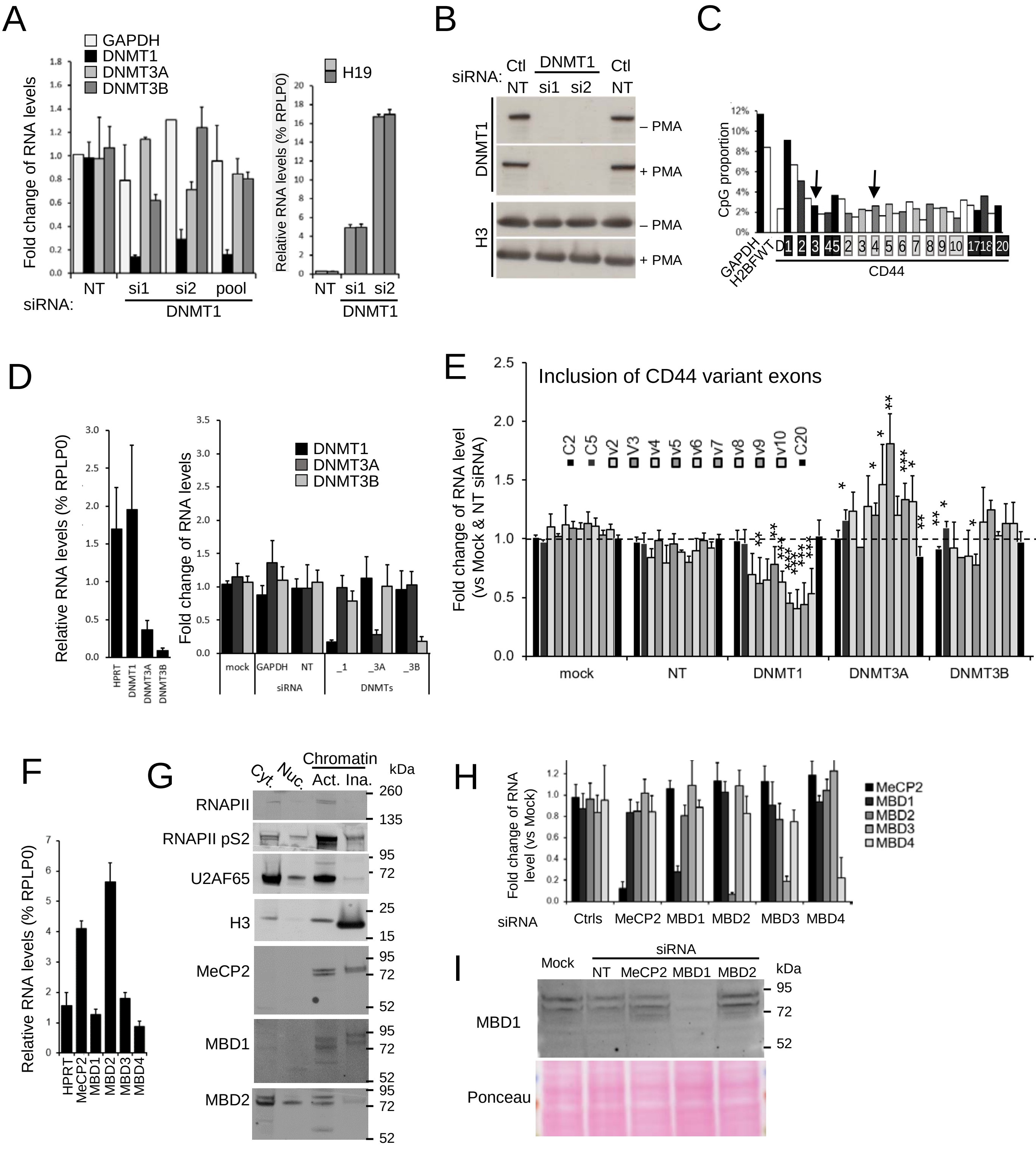

Batsché *et al.*, Suplementary figure S3 part1/2

J

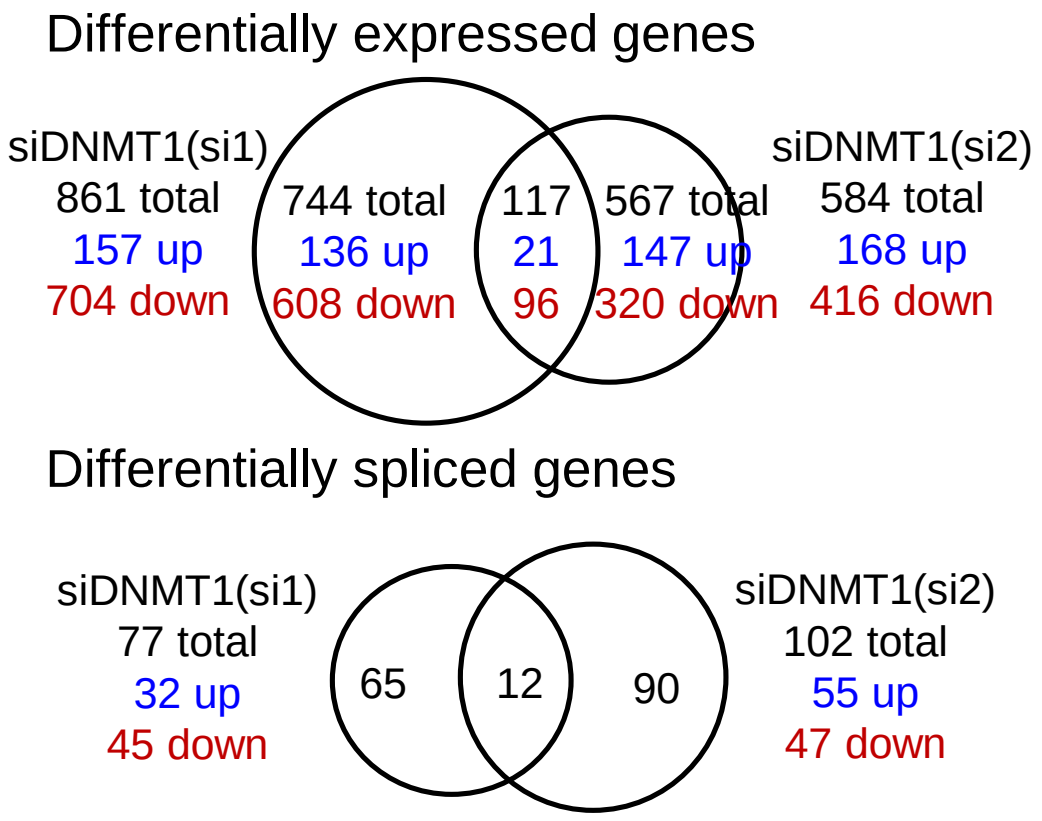

K

Commonly differentially spliced genes in both the 2 DNMT1 siRNAs after manual curation

AHRR e15 up (cassette exon) _ meCpG
DST ae39, e40 & e41/45 down (alt. Termination)_meCpG
GLS e18 to e20  up (e18=alt. prom) _ no CpG rich _ same than HP1
KIF1B  e28 to e54 up (e31=alt prom)_me CpG, e29 close CTCF paek
RABGAP1L e30 down / e32 to e37 up (e32=alt prom)_ CpG
CDRT1/TRIM16 ae5 to ae10 up (gene23ex)=alt start & end_ no CpG rich
GNAS e5e6 up / e9 to ae10 down (alt promoter ?)
CD44 e6 to e14 down (cassette exons) – no CpG rich

More strongly in si1 than in si2
NOTCH2NL e4e5e6 down / gene up
TXNIP ae2 down / gene up
MRPL11 alt prom e2 up / gene down
DAB1 e10e11 down / gene up

L

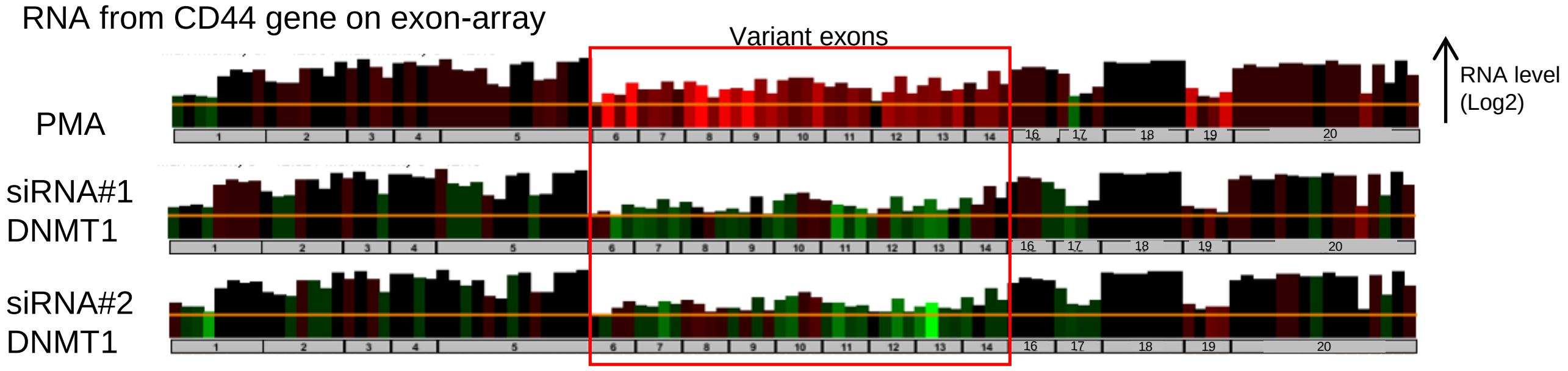

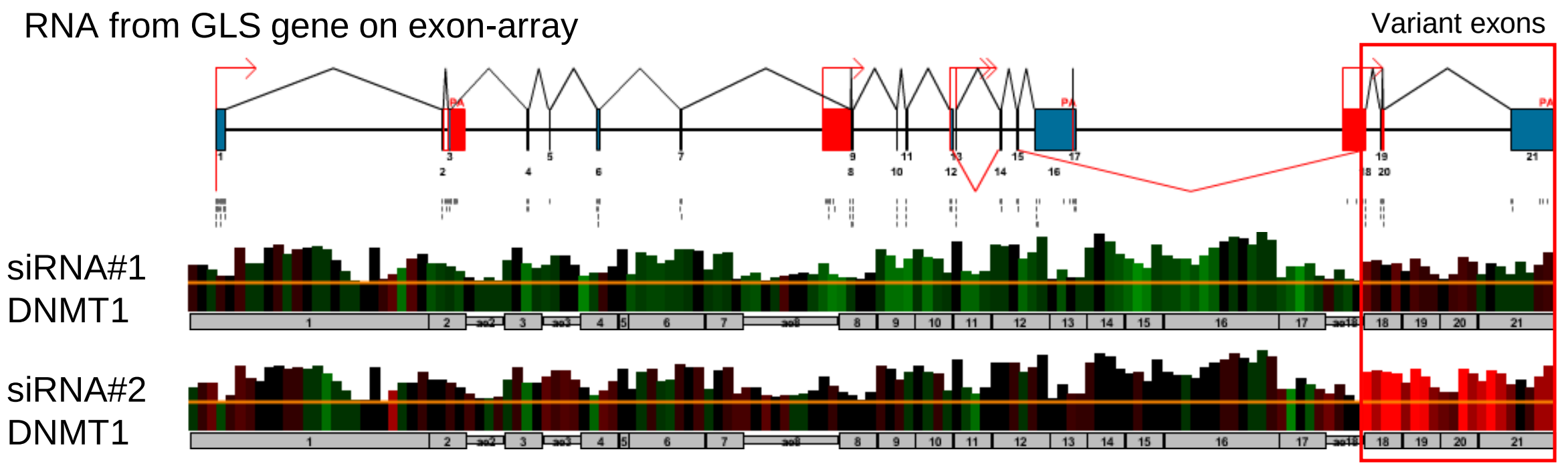

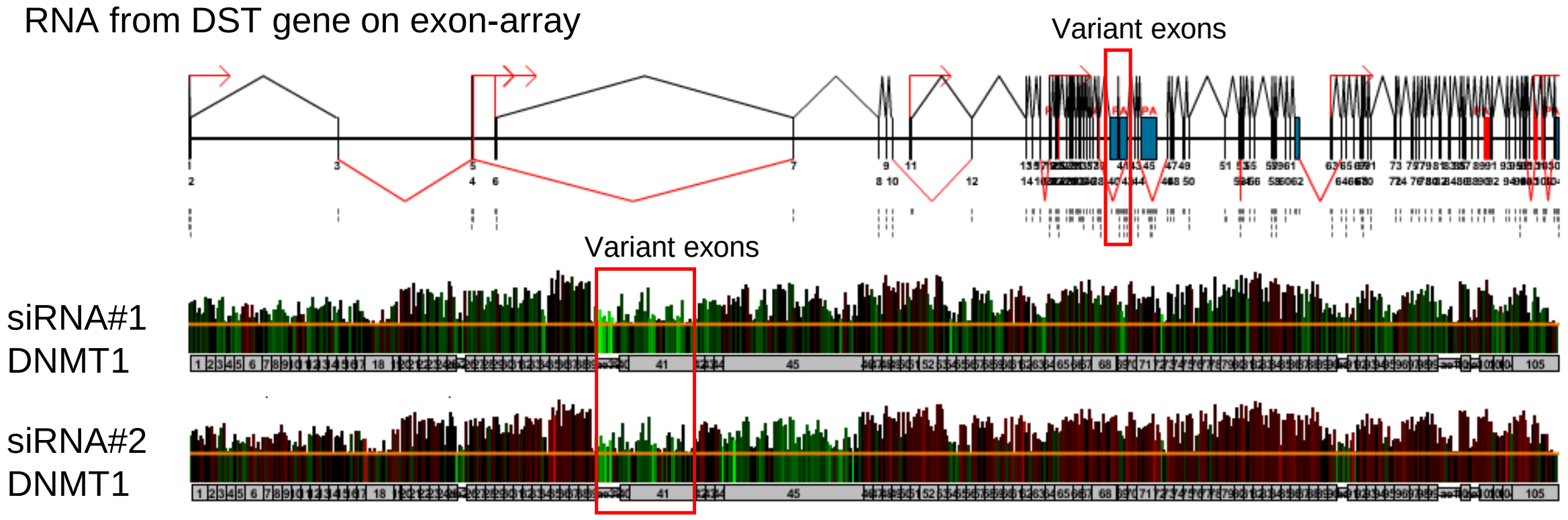

Batsché *et al.*, Suplementary figure S3 part2/2

**Supplemental figure S4**. **Analysis of expression pattern of Acute Leukemia (ALL).**
**A**) Principal component analysis of gene expression based on the 500 most variable genes evaluated by DESeq2 with Rlog normalization.
**B**) Proportion of reads aligning onto the hg19 human genome for each indicated library. The libraries with less than 60% of alignment were not considered for further analysis (white bars).
**C**) Differential gene expression between ALL and HCB cells evaluated by DESeq2 with a log2 fold change >1 and adjusted p-value <0.001 on 8 HCB and the 12 ALL samples which aligned equivalently onto the genome. Note that the genes from ChrX and ChrY were excluded from this analysis due to their absence in the SRR data of ALL samples from Almamun et al., (Almamun et al., 2015). 13244 genes were found expressed considering that their level exceeds the threshold defined at log2(cpm)=0.7 which correspond approx. to a minimum of 32 raw count per genes. Differential splicing of the genes evaluated by MAJIQ with high confidence P(|dPSI|>0.2)>0.95. The number of predicted genes having at least one alternative spliced event is indicated as well as those that are regulated downward (grey partition) or upward (black partition).
**D**) The CD44 expression level in each sample was calculated by averaging the reads covering the junction of constant exons. The data were presented as averages of exon-exon junctions (± s.e.m.). To analyze the inclusion of CD44 variant exons (Fig. 4A and 4B), the scared samples were used including the 8 normal samples (HCB) and the 9 ALL samples showing comparable range of CD44 global expression. Significant differences with T-Student p<0.05 (one-tail, equal var.) of CD44 alternative splicing were still found if only the 5 ALL expressing the highly CD44 level in the range of HCB (i.e. A20, A30, A37, A31, A17) were considered for comparison with the HCB.
**E**) The relative RNA levels of the indicated genes in the cohort of ALL patient versus the control cells. The RNA expression of indicated genes were evaluated by the normalized count of reads (cpm) covering the exons of the 12 selected ALL (Sup. fig. S5A) versus the 8 HCB. The indicated *P* value were calculated using the Student's t-test (two-tails) on the Rlog transformed read counts. Fold change (FC) were calculated on the median. RPLP0 reference gene is shown.
**F**) The separated tracks corresponding to the MIRA-seq from 18 ALL (blue) and 20 pre-B HCB (red), set at the same range (0-30), which are used for the overlay in the figure 4C.

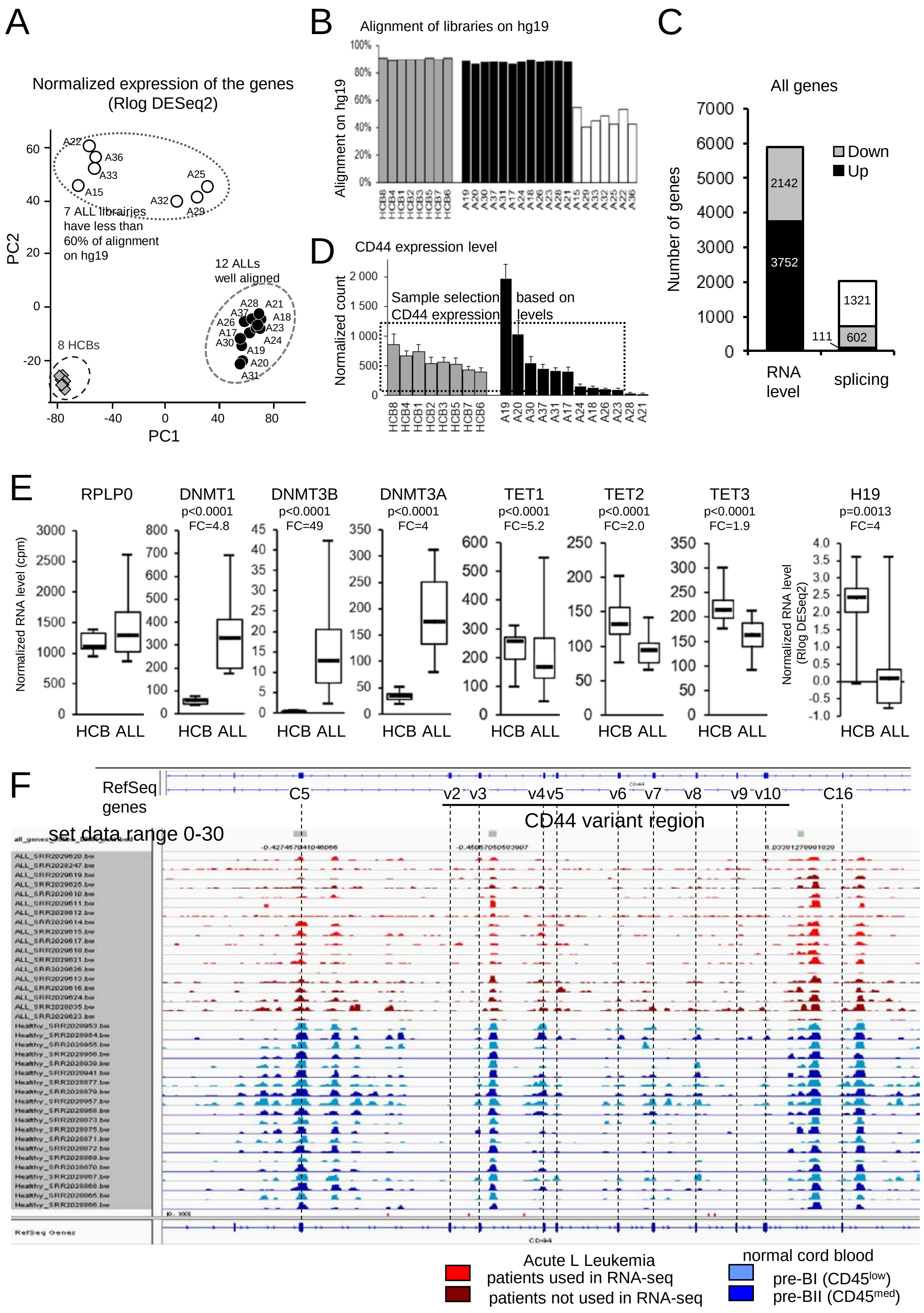

Batsché *et al.*, Suplementary figure S4

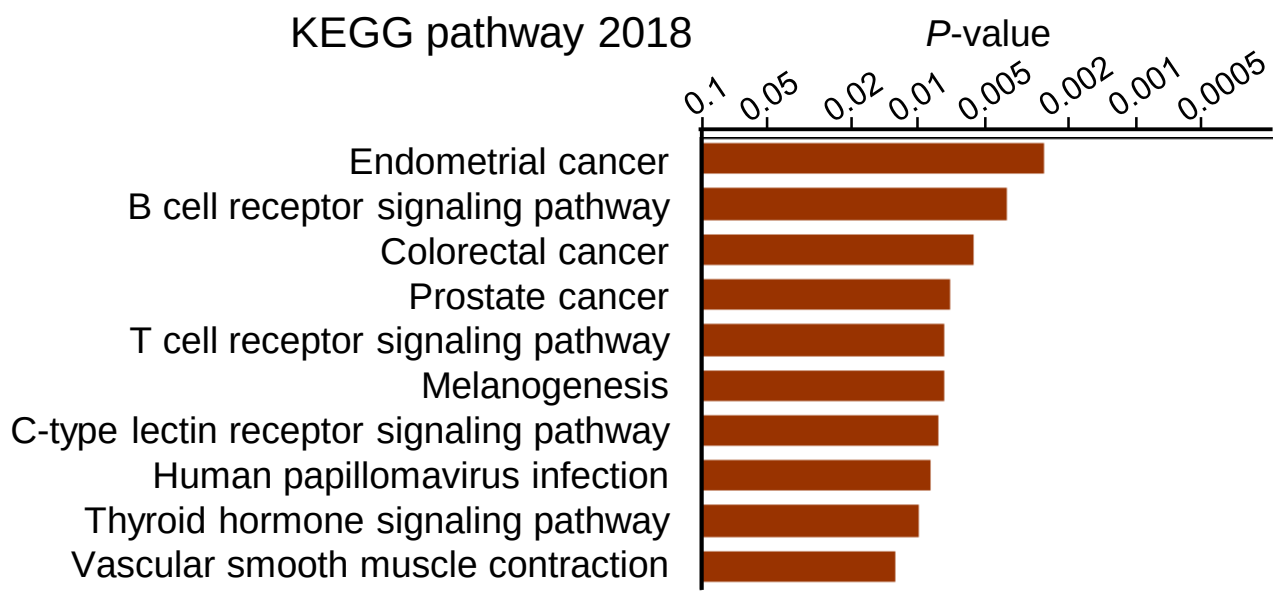

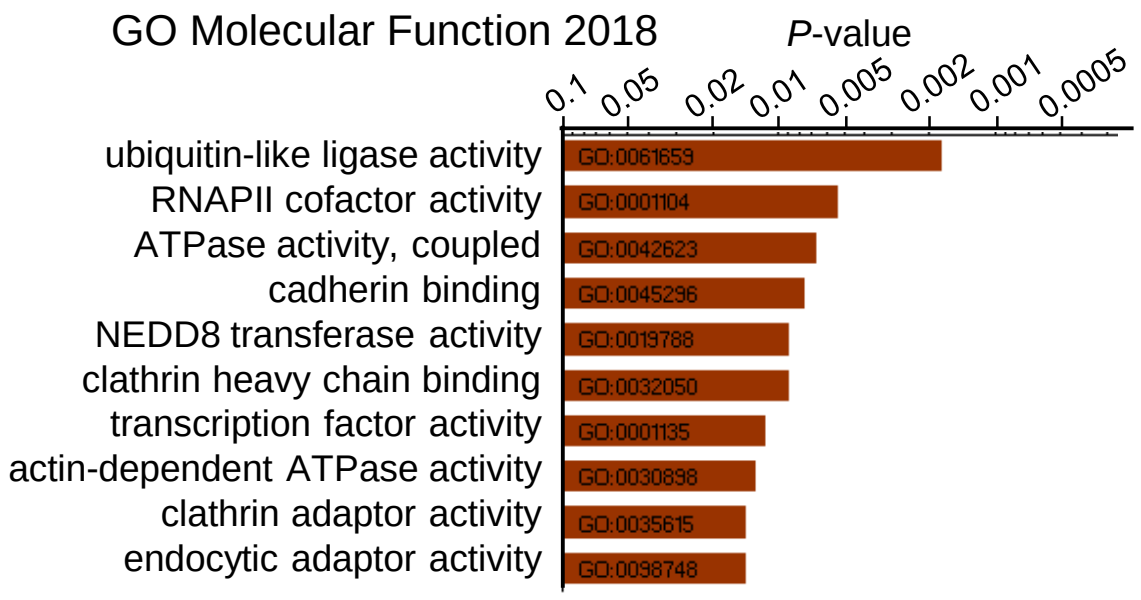

**Supplemental figure S5**. **pathway analyses of genes differentially spliced both in DKO and in ALL but not correlated with variation of local DNA methylation.**
Pathway analysis was performed using the Enrichr tool (Kuleshov et al., 2016) and the 26 genes with splice changes were not correlated with changes in DNA methylation (yellow, orange and khaki quarters of the figure 5A).

Batsché *et al.*, Suplementary figure S5